\documentclass[runningheads,anonymous,UKenglish, autoref, thm-restate]{elsarticle}
\usepackage[T1]{fontenc}
\usepackage[utf8]{inputenc}
\usepackage{graphicx}
\usepackage{natbib}

\usepackage{quoting}

\usepackage{environ}
\usepackage{xcolor}
\usepackage{array}
\usepackage{tikz}
\usepackage{colortbl}
\usepackage{multirow}
\usepackage{todonotes}
\usepackage[appendix=inline]{apxproof}
\usepackage{subcaption}
\usepackage[labelsep=quad, skip=3pt]{caption}
\usepackage{graphicx}

\usetikzlibrary{calc,shapes,patterns,positioning,fit,backgrounds,decorations,chains}
\usetikzlibrary{decorations.pathmorphing, shapes}
\usetikzlibrary{arrows}
\usetikzlibrary{decorations.markings}
\usetikzlibrary{arrows.meta}
\usetikzlibrary{chains,fit,shapes, shapes.multipart}
\usetikzlibrary{decorations.pathreplacing}
\usetikzlibrary{decorations.text, positioning,quotes}

\usepackage[linesnumbered,lined,boxed,ruled,noend]{algorithm2e}
\usepackage{amsmath,amssymb,amsfonts}
\usepackage{mathtools}
\usepackage{amsthm,nicefrac}
\usepackage[utf8]{inputenc}
\usepackage[normalem]{ulem}

\usepackage{float}
 \usepackage{amssymb}
\usepackage{pifont}

\usepackage{hyperref}
 \usepackage{cleveref}
 \crefname{subsection}{subsection}{subsections}
 \Crefname{subfigure}{Figure}{Figures}

\newcommand{\prob}[3]
{
\begin{quote}
  {\sc #1}
  \nopagebreak
  \begin{description}
  \item[\textbf{Input:}] #2
  \item[\textbf{Question:}] #3
  \end{description}
\end{quote}
}

\tikzstyle{smallvertex}=[circle, draw, inner sep=2pt]
\tikzstyle{rectvertex}=[rectangle, draw, inner sep=4pt]
\tikzstyle{doublevertex}=[circle, draw, inner sep=2pt,double]
\tikzstyle{m}=[line width=1mm]

\definecolor{goldenpoppy}{rgb}{0.99, 0.76, 0.0}
\definecolor{iris}{rgb}{0.35, 0.31, 0.81}
\definecolor{forestgreen}{rgb}{0.13, 0.55, 0.13}

\newtheorem*{lemma*}{Lemma} 

\newtoggle{sketch}
\togglefalse{sketch}
\makeatletter
 \@ifpackagewith{apxproof}{appendix=inline}{}{}
 \ifthenelse{\equal{\axp@appendix}{strip}}{}{}
 \NewEnviron{sketch}
 { \ifthenelse{\equal{\axp@appendix}{inline}}{}{
     \iftoggle{sketch}{
     \begin{proofsketch}
       \BODY
     \end{proofsketch}
       }{}
   }}
 \makeatother

\newtheoremrep{theorem}[ctheorem]{Theorem}
 \Crefname{ctheorem}{Theorem}{Theorems}

\newtheoremrep{lemma}[clemma]{Lemma}
 \Crefname{clemma}{Lemma}{Lemmas}

\newtheoremrep{corollary}[ccorollary]{Corollary}

\newtheoremrep{observation}[cobservation]{Observation}
\Crefname{observation}{Observation}{Observations}
\newcounter{cexample}
\newtheorem{exampleB}[cexample]{Example}

\newtheorem{construction}{Construction}

\newtheorem{definition}{Definition}
\makeatletter
\newcommand{\letter}[1]{\@alph{#1}}
\newcommand{\Letter}[1]{\@Alph{#1}}
\makeatother

\journal{Discrete Applied Mathematics}
\begin{document}

\begin{frontmatter}

\title{Parameterised algorithms for temporally satisfying reconfiguration problems}
\author[label1,label2]{Tom Davot}
\ead{tom.davot@univ-angers.fr}
 % \orcidID{0000-0003-4203-5140}
\author[label2]{Jessica Enright}
\ead{jessica.enright@glasgow.ac.uk}
% \orcidID{0000-0002-0266-3292}\,\email{jessica.enright@glasgow.ac.uk}
\author[label2]{Laura Larios-Jones\corref{cor1}}
\ead{l.larios-jones.1@research.gla.ac.uk}
\cortext[cor1]{Corresponding author}
% \,\orcidID{0000-0003-3322-0176}\,\email{l.larios-jones.1@research.gla.ac.uk}
\affiliation[label1]{organization ={Univ Angers, LERIA, SFR MATHSTIC}, addressline={F-49000 Angers}, country={France}}
\affiliation[label2]{organization ={School of Computing Science, University of Glasgow}, country={UK}}

 % \nolinenumbers %uncomment to disable line numbering

% % typeset the header of the contribution
%

\begin{abstract}
Given a static vertex-selection problem (e.g. independent set, dominating set) on a graph, we can define a corresponding temporally satisfying reconfiguration problem on a temporal graph which asks for a sequence of solutions to the vertex-selection problem at each time such that we can reconfigure from one solution to the next. We can think of each solution in the sequence as a set of vertices with tokens placed on them; our reconfiguration model allows us to slide tokens along active edges of a temporal graph at each time-step.  

We show that it is possible to efficiently check whether one solution can be reconfigured to another, and show that approximation results on the static vertex-selection problem can be adapted with a lifetime factor to the reconfiguration version. Our main contributions are fixed-parameter tractable algorithms with respect to: enumeration time of the related static problem; the combination of temporal neighbourhood diversity and lifetime of the input temporal graph; and the combination of lifetime and treewidth of the footprint graph.

\end{abstract}
% \begin{nolinenumbers}
%     \input{changes-made}
% \end{nolinenumbers}

% \begin{highlights}
% \item Introduction of temporally satisfying reconfiguration problems.
% \item Proof of tractability with respect to
% \begin{itemize}
%     \item enumeration time of the related static problem,
%     \item temporal neighbourhood diversity and lifetime,
%     \item treewidth of the footprint graph and lifetime by Courcelle's theorem,
% \end{itemize}
% in general, and applications of these results to \textsc{Temporal Dominating Set Reconfiguration} and \textsc{Temporal Independent Set Reconfiguration}
% \end{highlights}
\end{frontmatter}

\section{Introduction}

In many classical graph problems in static settings, the aim is to select a subset of vertices respecting some specified property, while minimizing or maximizing the number of selected vertices (e.g.\ maximum independent set, minimum dominating set, maximum clique).
Recently, reconfiguration versions of classical vertex-selection problems have been studied on static graphs, where the goal is typically to determine whether one valid solution can be transformed into another through a sequence of small modifications~\cite{BousquetMNS24,MouawadN0SS17}.

Formally, reconfiguration versions of vertex-selection problems in static graphs can be defined as follows. Let $\Pi$ be a vertex-selection problem such as {\sc Independent Set} or {\sc Dominating Set}. Reconfiguration problems on static graphs are often defined with a starting and end solution $S$ and $T$ respectively. The aim is to determine if there is a sequence of vertex sets starting with $S$ and ending with $T$ such that each is a solution for $\Pi$ and consecutive sets are adjacent in the space of solutions. The definition of solution adjacency varies according to model. In the \emph{token sliding model}, a solution can be seen as a set of tokens placed on the vertices of $G$~\cite{HEARN200572}. A solution $S_1$ is adjacent to another solution $S_2$ if we can transform $S_1$ into $S_2$ by sliding one token of $S_1$ along an edge $uv$ of $G$ such that $S_1 \setminus S_2 = \{u\}$ and $S_2 \setminus S_1 = \{v\}$. Both {\sc Independent Set Reconfiguration} and {\sc Dominating Set Reconfiguration} have been studied in parameterised~\cite{BodlaenderGS21,BartierBDLM21} and classical~\cite{BousquetJO20,SuzukiMN16} settings. 
% Other models, such as \emph{token jumping model}, and other reconfiguration problems have been studied, see~\cite{BousquetMNS24,MouawadN0SS17} for an overview.

Hearn and Demain~\cite{HEARN200572} show that {\sc Independent Set Reconfiguration} under token sliding is PSPACE-complete even in planar graphs with bounded degree. It is also known to be PSPACE-complete when restricted to other classes of graphs such as split graphs~\cite{BelmonteKLMOS21}, bipartite graphs~\cite{LokshtanovM19}, and graphs of bounded bandwidth~\cite{Wrochna18}. On the positive side, there are polynomial time algorithms when restricted to trees~\cite{DemaineDFHIOOUY14}, interval graphs~\cite{BonamyB17}, bipartite permutation graphs~\cite{Fox-EpsteinHOU15} and line graphs~\cite{ItoDHPSUU11}. 

For {\sc Dominating Set Reconfiguration} under token sliding, the problem is also known to be PSPACE-complete when restricted to split, bipartite and bounded tree-width graphs~\cite{HaddadanIMNOST16}, or restricted to circle graphs~\cite{BousquetJ21}. The problem is polynomial-time solvable on dually chordal graphs, cographs~\cite{HaddadanIMNOST16} and on circular-arc graphs~\cite{BousquetJ21}.
% Some structural properties have also been studied .
 
These reconfiguration problems provide valuable insights into how dynamic solutions can evolve in static graphs, but they cannot model scenarios where the underlying structure of the graph evolves.
While reconfiguration on temporal graphs has already been studied before (see, for example, {\sc Temporal Arborescence reconfiguration}~\cite{ito2023reconfiguration,dondi2024complexity}), we propose a natural variation on the common reconfiguration setup whereby reconfiguration steps coincide with timesteps of a temporal graph.
Based on some vertex selection problem, we define a \emph{temporally satisfying reconfiguration problem}, which differs slightly from the common reconfiguration setup: instead of being given a starting and end solution, we instead ask if there exists a sequence of solutions for each time-step such that each set can be reconfigured into the next via token sliding. Moreover, while the token sliding model in static settings only allows one token to move at a time, we generalise by allowing all tokens to move along available edges simultaneously at each time-step.
Verheije studied a version of this problem for {\sc Dominating Set} on temporal graphs~\cite{Verheije21}, under the name {\sc Marching Dominating}. He developed an exact exponential-time algorithm to determine whether a solution of {\sc Dominating Set} with at most $k$ tokens can be maintained in a temporal graph. 
Importantly, in our model, we require the vertices selected at any time to be a solution to the vertex-selection problem on the snapshot at that time. This is what gives these reconfiguration problems the name ``temporally satisfying''. We also note that all of our results hold on the restrictions of the problems where the start and end solution are given -- as is typically the case in static reconfiguration problems.

%In a similar vein the problem {\sc Temporal Arborescence Reconfiguration} gives a starting and finishing temporal arborescence (a subset of temporal arcs in a directed temporal graph which induce a temporal subgraph such that there is a unique directed temporal path from a root to each vertex) and looks for a sequence of reconfigurations between the arborescences such that each intermediate set of temporal arcs forms a temporal arborescence. Here the reconfiguration operation on the arcs allows an arc to be ``flipped'', meaning it is replaced with an arc which has not been selected. If there is no bound on the number of reconfigurations which may be performed, the problem is known to be solvable in polynomial time~\cite{ito2023reconfiguration}, but has been shown to be W[1]-hard with respect to number of reconfigurations plus lifetime of the temporal graph~\cite{dondi2024complexity}. 

The paper is organised as follows: in Section~\ref{sec:notation}, we introduce basic graph and temporal graph notation.  In \Cref{sec:tmp reconfiguration problems}, we formally introduce temporally satisfying reconfiguration problems and define a transformation of any vertex-selection problem on a static graph into a temporally satisfying reconfiguration problem. In \Cref{sec:polynomial}, we describe two preliminary results on temporal graphs, showing we can check if a specified sequence is temporally reconfigurable, and giving an approximation result. In Sections~\ref{sec:enum},~\ref{sec:ndiversity}, and~\ref{sec:tw algo} we give parameterised algorithmic results: in \Cref{sec:enum}, we study the parameterisation of temporally satisfying reconfiguration problems by the enumeration time of their static versions; in \Cref{sec:ndiversity}, we parameterise by temporal neighbourhood diversity and lifetime; finally, in \Cref{sec:tw algo}, we parameterise by treewidth of the footprint graph. 
%%% Local Variables:
%%% mode: LaTeX
%%% TeX-master: "main"
%%% End:

\section{Notation}
\label{sec:notation}
Where possible, we use standard graph theoretic notation, and direct the reader to \cite{golumbic} for detail.  A static graph is a pair $G = (V, E)$ where $V(G) = V$ is the vertex set and $E(G) = E$ is the edge set. We also use standard definitions and notation of parameterised algorithmics as in~\cite{cygan_parameterized_2015}.
A \emph{matching} $M \subseteq E(G)$ of graph $G$ is a set of edges that are pairwise disjoint. A matching $M$ is \emph{perfect} if each vertex $v \in V(G)$ is in exactly one edge in $M$.  Finally, given two disjoint subsets of vertices $X$ and $Y$, we say that there is a \emph{perfect matching between} $X$ and $Y$ if there exists a perfect matching in the bipartite subgraph $H$ with $V(H) = X \cup Y$ and $E(H) = \{xy \mid xy \in E(G), x \in X, y\in Y \}$. 

% I think we can probably get away without defining the below.
% A \emph{mapping function} $f : D \to 2^I$ is a function that associates some elements of $D$ to a (possibly empty) subset of $I$. We usually use the functional notation to describe a mapping function, that is, $f(x)=S$ indicates that $x\in D$ is associated to the subset $S \in I$. For conveniency, we also can see $f$ as a set of relations $x \to y$, where $x$ and $y$ are elements of $D$ and $I$, respectively. Hence, we have $f(x)= \{ y\in I \mid x \to y \in f\}$. It is particulary convenient when constructing a mapping function $f_2$ from another mapping function $f_1$: for example if $f_2 = f_1 \cup \{x_1 \to y_1\}$, then we have $f_2(x)=f_1(x) \cup \{y_1\}$ if $x=x_1$ and $f_2(x)=f_1(x)$ otherwise. 
%

\subsection{Temporal graphs}
A \emph{temporal graph} $\mathcal{G}$ is a pair $(G, \lambda)$, where $G$ is a static graph and $\lambda : E(G) \to 2^{\mathbb{N}}$ is a function called the \emph{time-labelling function}; for each edge $e \in E(G)$, $\lambda(e)$ denotes the set of time-steps at which $e$ is active. 
The \emph{lifetime} of $\mathcal{G}$, denoted by $\tau_{\mathcal{G}}$, is the maximum time-step at which an edge is active, i.e., $\tau_{\mathcal{G}} = \max \bigcup_{e \in E(G)} \lambda(e)$. 
When $\mathcal{G}$ is clear from the context, we may drop the subscript and simply write $\tau$. 
For a temporal graph $\mathcal{G} = (G, \lambda)$ and $t \in [\tau]$, the \emph{snapshot} of $\mathcal{G}$ at time-step $t$ is the static graph $G_t$ that consists of all the edges of $G$ that are active at time-step $t$, i.e.,  $V(G_t) = V(G)$ and $E(G_t) = \{e \in E(G) \mid t \in \lambda(e)\}$. For $v \in V(G_t)$, we use $N_t(v)$ to denote the set of neighbours of $v$ in the graph $G_t$. That is, $N_t(v)=\{u\in V(G) : uv\in E(G_t)\}$. When $v$ is included in the set of neighbours, we use the notation $N_t[v]$ and refer to the vertices in the set as the closed neighbours of $v$.
The \emph{footprint} of a temporal graph is the static graph formed by taking the union of the temporal graph at all time-steps.

\section{Temporally satisfying reconfiguration problems}
\label{sec:tmp reconfiguration problems}
We now define temporally satisfying reconfiguration problems and give some intuition.  Here we define a temporally satisfying reconfiguration problem from a static graph vertex selection problem: any optimisation problem in which we choose a set of vertices that respect some property, with the aim of either maximising or minimising the number of vertices in the selected set.  
Intuitively, we can think of a set of selected vertices as a set of tokens placed on those vertices.  
Then, we require the vertices indicated by tokens to meet a required property at each time-step, and allow movement over an edge of each token at every time-step.
Importantly, each vertex can only contain at most one token. At each time-step and for each token, we have two possibilities: either to keep the token on the same vertex or to move the token to an adjacent vertex.  We then ask for the optimum number of tokens such that we respect the required property at each time and can also reconfigure the tokens from each time-step to the next.  

Formally, 
let $T_1,T_2 \subseteq V(G)$ be two sets of selected vertices in a temporal graph $\mathcal{G}$. We say that $T_1$ is \emph{reconfigurable} into $T_2$ at time-step $t$ if it is possible to move the tokens from the vertices of $T_1$ to the vertices of $T_2$ in $G_t$. Observe that $T_1$ is reconfigurable into $T_2$ at time-step $t$ if there is a bijection $b : T_1 \to T_2$, called a \emph{reconfiguration bijection}, such that for each $(u,v) \in b$, either $u=v$ (i.e.\ the token remains on the same vertex) or $uv \in G_t$ (i.e.\ the token moves from $u$ to $v$ on the edge $uv$).

Notice that if $T_1$ is reconfigurable into $T_2$, then we have $|T_1|=|T_2|$. Let $T=(T_1,\dots,T_\tau)$ be a sequence such that for all $i \in [\tau]$, $T_i \subseteq V(G)$.  We say that $T$ is a \emph{reconfigurable sequence} of $\mathcal{G}$ if for each time-step $t \in [\tau-1]$, $T_t$ is reconfigurable into $T_{t+1}$ at time-step $t$. We denote by $|T|=|T_1|=\dots=|T_\tau|$ the \emph{size} of the sets in the reconfigurable sequence $T$.
In a temporally satisfying reconfiguration problem, we aim to find a reconfigurable sequence such that at each time-step $t$, the set of selected vertices $T_t$ respects a specific property $\Pi$ in $G_t$. This gives the formal basis for transforming a static vertex selection problem to a temporally satisfying reconfiguration problem. Consider the following generic vertex selection graph problem in a static setting:

 \prob{$\Pi$ static graph problem}{
   A static graph $G$.
}{
  Find a minimum/maximum set of vertices $X$ that respects property $\Pi$ in $G$.
}

\noindent The corresponding temporally satisfying reconfiguration problem can be formulated as follow:

 \prob{Temporal $\Pi$ reconfiguration problem}{
   A temporal graph $\mathcal{G}=(G,\lambda)$ with lifetime $\tau$.
 }{
   Find a reconfigurable sequence $T=(T_1,\dots,T_\tau)$ of $\mathcal{G}$ of minimum/maximum size $|T|$ such that each $T_t$ respects property $\Pi$ in $G_t$.
 }
 % \TD{I think it's better to keep the optimisation formulation if we want to keep \Cref{th:approx}}
Note that a reconfigurable sequence may be an optimal solution to a \textsc{Temporal $\Pi$ reconfiguration problem} without any of the sets being an \emph{optimal} solution to the static problem in the corresponding snapshot. To differentiate, we use solution to refer to optimal solutions of the static problems in question, and refer to a set as respecting the property $\Pi$ if the set is not necessarily optimal in cardinality.

 For the sake of simplicity, we refer to ``static problem'' to exclusively refer to a vertex selection problem in a static graph while ``temporally satisfying reconfiguration problem'' will refer to a temporally satisfying reconfiguration version of a static problem. 
 Notice that, since any static graph is a temporal graph of lifetime $1$, if a static problem is NP-hard, then the corresponding temporally satisfying reconfiguration problem is also NP-hard. More generally, any lower bounds on complexity can be transferred from the static problem to the temporally satisfying reconfiguration problem.
 As an example, we will use two classical problems, {\sc Dominating Set} and {\sc Independent Set} to illustrate the methods used throughout the paper. 

 \begin{exampleB}[Dominating Set]
   A \emph{dominating set} $D \subseteq V(G)$ is a set of vertices such that for each $v \in V(G)$, we have $N[v] \cap D \neq \emptyset$.
 \end{exampleB}

 A classical optimisation problem is to find a dominating set of minimum size in a static graph. This problem is known to be NP-complete~\cite{Garey90}.

 % \textcolor{magenta}{\textbf{Laura: max-edge result uses optimisation version of the problem. Below is also optimisation version, but title makes me think it was intended to be decision?}}

 % \prob{Temporal Reconfiguration Dominating Set}{ 
 %   A temporal graph $\mathcal{G}=(G,\lambda)$ with lifetime $\tau$.
 % }{
 %   Find a minimum-size reconfigurable sequence $T=(T_1,\dots,T_\tau)$ such that each $T_i$ is a dominating set.
 % }

\begin{exampleB}[Independent Set]
An \emph{independent set} $I \subseteq V(G)$ is a set of pairwise non-adjacent vertices.
\end{exampleB}
 
\noindent Finding an independent set of maximum size is one of Karp's NP-complete problems~\cite{Karp1972}.
In the following, we refer to the temporally satisfying reconfiguration versions of {\sc Dominating Set} and {\sc Independent Set} as {\sc Temporal Dominating Set Reconfiguration} and {\sc Temporal Independent Set Reconfiguration}, respectively.
\section{Preliminary results}
\label{sec:polynomial}
In the remainder of this paper we present a number of algorithmic results; here we start with several initial results first showing that we can efficiently check if a given sequence is reconfigurable, and then presenting a condition for a reconfiguration problem to belong to the same approximation class as its static version.

\subsection{Checking if a sequence is reconfigurable}

We show that testing if a sequence $(T_1,\dots,T_\tau)$ is a reconfigurable sequence of a temporal graph $\mathcal{G}=(G,\lambda)$ can be done in polynomial time. We begin with the smaller problem of checking whether one state can be reconfigured into another. This can be done by computing a perfect matching in a bipartite graph that captures what set changes are possible between time-steps. An example of this can be found in Figure~\ref{fig:reconfig-eg}. 

    \begin{figure}[ht]
  \centering
  \scalebox{0.85}{
    \begin{tikzpicture}[
tap/.style args = {#1}{decoration={raise=5,
                                      text along path,
                                      text align={align=center},
                                      text={#1}
                                      },
              postaction={decorate},
              },
                   ]

     \node[rectvertex,label=90:{$a$}] (a) at (-3,1) {};
     \node[smallvertex] at (-3,-1) {};
     % \foreach \a in {11,21,31}
     % \draw[->,tap={1/0/${+\infty}${}}] (s) -- node[midway,above] {} (x\a);

     \node[doublevertex,label=90:{$b$}] (b) at (-1,1) {};
     \node[rectvertex] at (-1,1) {};
     \node[doublevertex,label=-90:{$c$}] (c) at (-3,-1) {};
     \node[rectvertex] at (-3,-1) {};
     \node[doublevertex,label=-90:{$d$}] (d) at (-1,-1) {};
     \draw (a) -- (c);
     \draw (b) -- (c);
     \draw (d) -- (c);

    \node[smallvertex, label = 180 :{$v_a$}] (va) at (3,1.5) {};
    \node[rectvertex] at (3,1.5) {};
    \node[smallvertex, label = 180 :{$v_b$}] (vb) at (3,0.5) {};
    \node[rectvertex] at (3,0.5) {};
    \node[smallvertex, label = 180 :{$v_c$}] (vc) at (3,-0.5) {};
    \node[rectvertex] at (3,-0.5) {};
    \node[smallvertex, label = 180 :{$v_d$}] (vd) at (3,-1.5) {};

    \node[smallvertex, label = 0 :{$u_a$}] (ua) at (5,1.5) {};
    \node[doublevertex, label = 0 :{$u_b$}] (ub) at (5,0.5) {};
    \node[doublevertex, label = 0 :{$u_c$}] (uc) at (5,-0.5) {};
    \node[doublevertex, label = 0 :{$u_d$}] (ud) at (5,-1.5) {};

    \draw[dashed] (va) -- (uc);
     \draw (vb) -- (uc);
     \draw (vd) -- (uc);
     \draw (vc) -- (ua);
     \draw (vc) -- (ub);
     \draw[dashed] (vc) -- (ud);
     \draw (va) -- (ua);
     \draw[dashed] (vb) -- (ub);
     \draw (vc) -- (uc);
     \draw (vd) -- (ud);
   \end{tikzpicture}
   }
   \caption{\label{fig:reconfig-eg} Left: An example graph $G$. Vertices highlighted with squares are in $T_1$, vertices highlighted with circles are in $T_2$. Right: The construction $H$. Vertices highlighted with squares are in $T_1$, vertices highlighted with circles are in $T_2$. Dashed edges form a matching between vertices in $T_1$ and $T_2$.}
 \end{figure}
\begin{lemmarep}
  \label{lemma:check reconfiguration two sets}
Let $G$ be a static graph and let $T_1$ and $T_2$ be two subsets of vertices. We can determine in $\mathcal{O}((|V(G)| + |E(G)|) \cdot \sqrt{|V(G)|})$ time if $T_1$ is reconfigurable into $T_2$.
\end{lemmarep}  
\begin{proof}
  We first construct a bipartite graph $H$ with vertex set $V(H) = \{v_i \mid x_i \in T_1\} \cup \{u_i \mid x_i \in T_2\}$ and edge set $E(H) = \{v_iu_i \mid x_i \in T_1\cap T_2\} \cup \{v_iu_j \mid x_i \in T_1, x_j\in T_2, x_ix_j \in E(G)\}$. An example of $H$ can be found in Figure~\ref{fig:reconfig-eg}.
  We show that there is a reconfiguration bijection between $T_1$ and $T_2$ if and only if there is there is a perfect matching in $H$.
  Let $f$ be a reconfiguration bijection between $T_1$ and $T_2$. Consider the set of edges $M = \{ v_iu_j \mid i\neq j, x_i \in T_1, f(x_i) = x_j\} \cup \{ v_iu_i \mid x_i \in T_1, f(x_i) = x_i\}$. If $f(x_i)=x_i$, then $x_i \in T_1 \cap T_2$ and by construction of $H$, the edge $v_iu_i$ belongs to $H$. If $f(x_i)=x_j$, then $v_i \in T_1$, $v_j \in T_2$ and $x_ix_j \in E(G)$ and again by construction of $H$, $v_iv_j$ belongs to $H$. Hence, $M \subseteq E(H)$ and since $f$ is a bijection between $T_1$ and $T_2$, $M$ is a perfect matching in $H$.
  Now let $M$ be a perfect matching in $H$. For each edge $v_iu_j \in M$ (possibly $i=j$), we set $f(x_i)=x_j$. Clearly, each vertex of $T_1$ has eaxctly one image and each vertex of $T_2$ has exactly one inverse image, thus $f$ is a bijection. For each $f(x_i)=x_j$ (with $i\neq j$), we have $v_iu_j \in M$ and by construction of $H$, there is an edge between $x_i$ and $x_j$ in $G$. Hence, $f$ is a reconfiguration bijection.
  Finally, the graph $H$ can be constructed in $\mathcal{O}(|E(G)|+|V(G)|)$ and a perfect matching can be computed in $\mathcal{O}((|E(H)|+|V(H)|) \cdot \sqrt{|V(H)|})$~\cite{Hopcroft1971ANA}. We obtain an overall complexity of $\mathcal{O}((|V(G)| + |E(G)|) \cdot \sqrt{|V(G)|})$.
\end{proof}

%We call a sequence $T=(T_1,\dots,T_\tau)$ in temporal graph $\mathcal{G}$ \emph{reconfigurable} if for every $1 \leq i  < \tau$, the set $T_i$ is reconfigurable in $\mathcal{G}$ to $T_{i+1}$.  
We now extend the previous result to show that testing if a sequence is reconfigurable can also be done in polynomial time.

\begin{corollaryrep}
   \label{cor:check reconfiguration sequence}
Let $\mathcal{G}=(G,\lambda)$ be a temporal graph. Let $T=(T_1,\dots,T_\tau)$ be a sequence such that for all $i \in [\tau]$, $T_i \subseteq V(G)$. We can determine whether $T$ is a reconfigurable sequence in $\mathcal{O}(\tau \cdot (|V(G)|+|E(G)|\cdot\sqrt{|V(G)|}))$ time.
\end{corollaryrep}

\begin{proof}
A sequence $(T_1,\dots,T_\tau)$ is reconfigurable if and only if for each $i \in [\tau-1]$, $T_i$ is reconfigurable into $T_{i+1}$ in $G_i$. Hence, by \Cref{lemma:check reconfiguration two sets}, we can test if $(T_1,\dots,T_\tau)$ is reconfigurable in $\mathcal{O}(\tau \cdot (|V(G)|+|E(G)|\cdot\sqrt{|V(G)|}))$.
\end{proof}

\subsection{Approximation}

We now turn our attention to the approximation of temporally satisfying reconfiguration problems. An optimisation problem belongs to the approximation class $f(n)$-APX if it is possible to approximate this problem in polynomial time with an $\mathcal{O}(f(n))$ approximation factor. Let $\Pi_S$ be a minimisation static problem, and $\Pi$ be the desired property of a selected set of vertices. An $f$-approximation algorithm for $\Pi_S$ is a polynomial time algorithm that, given a graph $G$, returns an approximate solution $X_{app}$ such that $|X_{app}|\leq f(G) \cdot |X_{opt}|$, where $X_{opt}$ is an optimal solution for $\Pi_S$ in $G$. 
 \begin{theoremrep}
   \label{th:approx}
Let $\Pi_S$ be a minimisation static graph problem such that for any set $S$ respecting $\Pi$ and any vertex $v$, $S \cup \{v\}$ also respects $\Pi$. Let $\Pi_T$ be the corresponding temporally satisfying reconfiguration version of $\Pi_S$. If $\Pi_S$ is $f$-approximable, then $\Pi_T$ is $\tau\cdot f$-approximable.
\end{theoremrep}

\begin{proof}
  Let $app_S$ be a polynomial-time approximation algorithm for $\Pi_S$.
  Let $\mathcal{G}=(G,\lambda)$ be a temporal graph. Let $X_{app} = \cup_{t\in [1,\tau]} app_S(G_t)$ be the union of every approximate solution of $\Pi_S$ in all snapshots of $\mathcal{G}$.
  We show that the algorithm $app_T$ that returns the reconfigurable sequence $T_{app} = (X_{app},\dots,X_{app})$ (i.e. $T_{app}$ always contains $X_{app}$ as set of selected vertices) is a $\tau \cdot f(G)$ approximation algorithm for $\Pi_T$.  
 Since the same set is selected in every timestep, $T_{app}$ is obviously a reconfigurable sequence for $\Pi_T$, thus we only need to show that it achieves the desired ratio.
  For each time-step $t \in [1,\tau]$, let $X^t_{opt}$ denote the optimal solution for $\Pi_S$ in $G_t$. Let $T_{opt}=(T_1,\dots,T_\tau)$ be an optimal reconfiguration sequence for $\Pi_T$ in $\mathcal{G}$. Notice that for each time-step $t$, we have $|X^t_{opt}| \leq |T_t|$.  For all $t$, we have $|app_S(G_t)| \leq f(|G_t|) \cdot |X_{opt}^t|$ and thus, $|app_S(G_t)| \leq f(|G_t|) \cdot |T_t|$. It follows from the fact that we have taken the union of all approximate solutions over the lifetime of the temporal graph that $|X_{app}| = |T_{app}| \leq \tau \cdot f(G) \cdot |T_{opt}|$. We can conclude that $app_T$ is a $\tau\cdot f(G)$ approximation algorithm.
\end{proof}

{\sc Dominating Set} is known to be Log-APX-complete~\cite{EscoffierP06}, \textit{i.e.} there is a polynomial-time $log$-approximation algorithm and there is no polynomial-time approximation algorithm with a constant ratio. Further, the property that a set is a dominating set is closed under addition of more vertices. Thus, we obtain the following result.

\begin{corollary}
There is a $(\tau \cdot \log)$-approximation algorithm for {\sc Temporal Dominating Set Reconfiguration}.
\end{corollary}

\section{Fixed-parameter tractability with respect to the enumeration time of the static version}
\label{sec:enum}
 
In this section, we study the enumeration complexity of temporally satisfying reconfiguration problems. Such problems are inherently at least as hard as their static version, since any valid reconfiguration sequence must consist of feasible solutions at each snapshot. Nevertheless, we show that the temporal setting introduces only a moderate overhead: if the static solutions can be enumerated efficiently, then so can the temporal reconfigurations, with only a polynomial factor in the size and lifetime of the temporal graph.
Observe that the number of feasible solutions is generally exponential. For instance, many vertex-subset problems admit up to $2^n$ solutions. However, numerous problems are known to be efficiently enumerable under specific structural restrictions or parameterisations. Our approach demonstrates that, in such cases, the efficiency of the static enumeration can be transferred almost directly to the temporal setting.
To illustrate this, we analyse the problem with respect to the parameter $|E|_{\max}$, the maximum number of edges in any snapshot. This choice is motivated by the fact that, in many applications, temporal graphs are globally dense but remain sparse at each individual time step.
Note that while some enumeration problems ask to enumerate only maximal/minimal solutions, here we are interested in enumerating all sets that are feasible solutions at each snapshot, regardless of maximality/minimality. Indeed, in a reconfigurable sequence $(T_1,\dots,T_k)$, each set $T_i$ is not necessarily maximal/minimal. Accordingly, we use the term solution throughout this section to refer to any such a feasible set.

% We show in this section that if the number of sets respecting a property $\Pi$ can be bounded by some function $f(k)$ for some parameter $k$, then the associated temporally satisfying reconfiguration problem is in FPT with respect to $f(G)$. We show the enumeration time of \textsc{Temporal Dominating Set Reconfiguration} to be a function of the maximum number of edges in any snapshot alone. Although enumeration time is often large, the size of the input to temporal problems is largely considered to be $O(n+m+\tau)$ where $n$ is the number of vertices, $m$ is the number of \emph{underlying} edges and $\tau$ is the lifetime. There exist temporal graph who have a dense footprint, but which are sparse in every snapshot. This motivates the use of these parameters. 

\begin{lemmarep}
  \label{lemma:enumeration}
  Let $\Pi_S$ be a static problem and let $\Pi_T$ be the temporally satisfying reconfiguration version of $\Pi_S$. Let $\mathcal{G}=(G,\lambda)$ be a temporal graph such that for all $t \in [\tau]$, all sets respecting $\Pi$ in $G_t$ for $\Pi_S$ can be enumerated in $\mathcal{O}(f(G))$ time. We can solve $\Pi_T$ in $\mathcal{O}(f(G)^2 \cdot \tau \cdot (|V(G)|+|E(G)|)\cdot\sqrt{|V(G)|})$ time.
\end{lemmarep}

\begin{proof}
  For all $t \in [\tau]$, let $X_t$ denote the set of sets respecting $\Pi$ in $G_t$. For each $t \in [\tau]$, define $Y_t \subseteq X_t$ such that for all $T_t \in Y_t$, there exists a reconfigurable sequence $(T_1,\dots,T_t)$ in the temporal graph containing the $t$ first snapshots of $\mathcal{G}$.
  Notice that $Y_\tau$ contains all sets $T_\tau$ such that there is a reconfigurable sequence ending with $T_\tau$ that is a solution for $\Pi_T$. Hence, to obtain the optimal size for $\Pi_T$ in $\mathcal{G}$, it suffices to take the size of the best (minimised or maximised as required by $\Pi_S$) set  $T_\tau \in Y_\tau$ for $\Pi_S$ in $G_\tau$.

  We now show by induction how we construct $Y_t$ for each $t \in [\tau]$.
  For the base case, we set $Y_1 = X_1$. Clearly all sets in $Y_1$ are sets respecting $\Pi$ which are in a trivial reconfigurable sequence.
  Now suppose, for induction, that $Y_{t-1}$ is the set of sets $T_{t-1}$ respecting $\Pi$ such that there exists a reconfigurable sequence $T_1,\ldots,T_{t-1}$ in the temporal graph containing the $t$ first snapshots of $\mathcal{G}$. We set $Y_t$ to be the subset of $X_t$ such that, for each $T_t\in Y_t$ there exists $T_{t-1} \in Y_{t-1}$ such that $T_{t-1}$ is reconfigurable into $T_t$ in $G_{t-1}$. Let $T_t \in Y_t$. First, since $T_t \in X_t$, $T_t$ respects $\Pi$ in $G_t$. Then, for each $T_t \in Y_t$, there is a $T_{t-1} \in Y_{t-1}$ such that $T_{t-1}$ is reconfigurable into $T_t$ in $G_{t-1}$. By the inductive hypothesis, there is a reconfigurable sequence $(T_1,\dots,T_{t-1})$ for $\Pi_T$ in the temporal graph containing the $t-1$ first snapshots of $\mathcal{G}$. Hence, we can conclude that the sequence $(T_1,\dots,T_{t-1},T_t)$ is a reconfigurable sequence for $\Pi_T$ in the temporal graph containing the $t$ first snapshots of $\mathcal{G}$. Hence, all sets in $Y_t$ respect the inductive hypothesis.
  Now let $(T_1,\dots,T_t)$ be a reconfigurable sequence for $\Pi_T$ in the temporal graph containing the first $t-1$ snapshots of $\mathcal{G}$. That is, $T_{t-1}$ is reconfigurable into $T_t$ in $G_t$ and $(T_1,\dots,T_{t-1})$ be a reconfigurable sequence for $\Pi_T$ in the temporal graph containing the $t-1$ first snapshots of $\mathcal{G}$. By the inductive hypothesis, $T_{t-1} \in Y_{t-1}$ and so, by construction of $Y_t$, $T_t$ belongs to $Y_t$. We can conclude that $Y_t$ respects the inductive hypothesis.
  
  We check reconfigurability for all pairs of sets in $X_t\times X_{t+1}$ for all $t\in [1,\tau)$. Since $|X_t| \in \mathcal{O}(f(G))$ for all $t$, by \Cref{lemma:check reconfiguration two sets} each $Y_t$ can be computed in $\mathcal{O}(f(G)^2\cdot (|V(G)| + |E(G)|)\cdot \sqrt{|V(G)|})$. Hence, computing $Y_\tau$ can be done in $\mathcal{O}(f(G)^2\cdot \tau \cdot (|V(G)| + |E(G))|\cdot \sqrt{|V(G)|})$. Notice that it is possible to associate each set $T_t$ with a sequence $(T_1,\dots,T_t)$ to make a constructive algorithm.
\end{proof}

% \subsection{FPT with respect to maximum number of edges in any snapshot}
% \label{sec:FPT with max edges}

Let $|E|_{\max}$ denote $\max_{t\in |\tau|}|E_t|$, the maximum number of edges in any snapshot of the temporal graph $\mathcal{G}$.
We use \Cref{lemma:enumeration} to show inclusion of \textsc{Temporal Dominating Set Reconfiguration} in FPT with respect to $|E|_{\max}$. First, we show that the number of dominating sets is bounded by a function of $|E|_{\max}$ in each snapshot.

\begin{lemmarep}
  Let $\mathcal{G}=(G,\lambda)$ be a temporal graph. For each $t \in [\tau]$, all dominating sets of $G_t$ can be enumerated in $\mathcal{O}(2^{2|E|_{max}} \cdot |E|_{max})$ time.
\end{lemmarep}

\begin{proof}
  Let $G_t$ be a snapshot of $\mathcal{G}$, $V_0$ be the set of vertices of degree zero in $G_t$ and $V_{\geq 0} = V(G) \setminus V_0$. 
  Notice that in any dominating set $D$ of $G_t$, $V_0 \subseteq D$. Hence, to enumerate all dominating sets in $G_t$, it suffices to list every dominating set in the subgraph induced by $V_{\geq 0}$.  For each enumerated dominating set $D$ for $G_t[V_{\geq 0}]$, a dominating set for $G_t$ can be obtained by taking the union of $V_0$ and $D$.
  To list all dominating sets $D$ for $G_t$, we enumerate all subsets of $V_{\geq 0}$ and check for each one whether it forms a dominating set. The enumeration of all subsets of $V_{\geq 0}$ can be done in $\mathcal{O}(2^{|V_{\geq 0}|})$ and checking whether a subset of $V_{\geq 0}$ is a dominating set can be done in $\mathcal{O}(|V_{\geq 0}| + |E|_{max})$.
  Since we have $|V_{\geq 0}| \leq 2|E|_{max}$, we can conclude that all dominating sets of $G_t$ can be enumerated in $\mathcal{O}(2^{2|E|_{max}} \cdot |E|_{max})$.
\end{proof}

\noindent Now, we can conclude that \textsc{Temporal Dominating Set Reconfiguration} is in FPT with respect to $|E|_{\max}$.

\begin{corollary}
{\sc Temporal Dominating Set Reconfiguration} can be solved in $\mathcal{O}(2^{4|E|_{max}} \cdot |E|_{max}^2 \cdot \tau \cdot(|V(G)|+|E(G)|\cdot\sqrt{|V(G)|}))$ time.
\end{corollary}

% \subsection{FPT with respect to minimum-size clique cover}

% Let $G$ be a static graph. A \emph{clique cover} $P = \{X_1,\dots,X_k\}$ of $G$ is a partition of its vertices into cliques. We 

% \begin{lemma}
% Let $\mathcal{G}=(G,\lambda)$ be a temporal graph such that there are two constants $k_1$ and $k_2$ such that for each $t \in [\tau]$, the vertices of $G_t$ can be partitionned into at most $k_1$ vertex-disjoint cliques of size at most $k_2$. For each $t \in [\tau]$, all independent sets of $G_t$ can be enumerated in $\mathcal{O}(k_2^{k_1} \cdot |E(G)|)$.
% \end{lemma}

% \begin{corollary}
% {\sc Reconfiguration Independent Set} can be solved in $\mathcal{O}(k_1^{k_2} \cdot \tau \cdot (|V(G)|+|E(G)|\cdot\sqrt{|V(G)|}))$.
% \end{corollary}

%%% Local Variables:
%%% mode: LaTeX
%%% TeX-master: "main"
%%% End:

\section{Fixed parameter tractability by lifetime and temporal neighbourhood diversity}
\label{sec:ndiversity}
\newcommand{\tndg}[1][\mathcal{G}]{\ensuremath{TND_{#1}}\xspace}
\newcommand{\ndg}[1][\mathcal{G}]{\ensuremath{ND_{#1}}\xspace}
In this section we present a fixed-parameter tractable algorithm, parameterised by lifetime and the temporal neighbourhood diversity of the temporal graph, to solve a class of temporally satisfying reconfiguration problems that we call \emph{temporal neighbourhood diversity locally decidable}. 

We need a number of algorithmic tools to build toward this overall result.  First, in \Cref{subsec:definitions}, we give definitions and notation necessary for this section, including defining temporal neighbourhood diversity, as well as the class of temporal neighbourhood diversity locally decidable problems. These build on analogous definitions in static graphs. 

%In \Cref{subsec:reconfiguration}, we introduce a key subroutine of the parameterized algorithm, which we then use in Subsection \Cref{subsec:algo}
Then, in \Cref{subsec:algo} we describe an overall algorithm to solve our restricted class of problems that is in FPT with respect to temporal neighbourhood diversity and lifetime. This algorithm uses a critical subroutine that constitutes the majority of the technical detail, and is presented in \Cref{subsec:reconfiguration}.  The subroutine uses a reduction to the efficiently-solvable circulation problem to give the key result (in \Cref{lemma:compute reconfigurable sequence}). The result allows us to efficiently generate an optimal reconfigurable sequence that is compatible with a candidate reconfiguration sequence of a type specific to temporal neighbourhood diversity locally decidable problems.

% Then, we present the intuition of the algorithm in \Cref{subsec:algo}. Finally, we describe and show the correctness of the main subroutine of the algorithm in \Cref{subsec:reconfiguration}.

\subsection{Definitions}
\label{subsec:definitions}

% \subsubsection{In static graphs}

Neighbourhood diversity is a static graph parameter
% We first state the formal definitions of neighbourhood diversity and neighbourhood diversity locally decidable problem. This parameter has been
originally introduced by Lampis~\cite{Lampis12}:

\begin{definition}[Neighbourhood Diversity~\cite{Lampis12}]
  The \emph{neighbourhood diversity} of a static graph $G$ is the minimum value $k$ such that the vertices of $G$ can be partitioned into $k$ classes $V_1,\dots,V_k$ such that 
  for each pair of vertices $u$ and $v$ in a class $V_i$, we have $N(u) \setminus \{v\} = N(v) \setminus \{u\}$.  We call $V_1,\dots,V_k$ a \emph{neighbourhood diversity partition} of ${G}$.
\end{definition}

Notice that each set $V_i$ of $P$ forms an independent set or a clique. Moreover, for any pair of sets $V_i$ and $V_j$ either no vertex of $V_i$ is adjacent to any vertex of $V_j$ or every vertex of $V_i$ is adjacent to every vertex of $V_j$. We distinguish two types of classes: $V_i$ is a \emph{clique-class} if $G[V_i]$ is a clique and $V_i$ is an \emph{independent-class} otherwise.
%
% Observe that the definition of temporal neighbourhood diversity is a generalisation of the neighbourhood diversity in static graphs since the temporal neighbourhood diversity partition of a temporal graph with lifetime one and the neighbourhood diversity partition of its footprint are the same. 

%

 \begin{definition}[Neighbourhood diversity graph]
 Let $G$ be a static graph with neighbourhood diversity partition $V_1,\dots,V_k$. The \emph{neighbourhood diversity graph} of $G$, denoted $\ndg[G]$ is the graph obtained by merging each class $V_i$ into a single vertex. Formally, we have $V(\ndg[G]) = \{V_1,\dots,V_k\}$ and $E(\ndg[G]) = \{V_iV_j \mid \forall v_i \in V_i, \forall v_j\in V_j, v_iv_j \in E(G)\}$.
\end{definition}

For clarity, we use the term class when referring to a vertex of \ndg[G], in order to distinguish between the vertices of $G$ and the vertices of \ndg[G].
%
% The definition of temporal neighbourhood diversity in temporal graphs is a generalisation of the neighbourhood diversity which is a parameter defined in static graphs. The neighbourhood diversity of a static graph $G$ is equivalent to the temporal neighbourhood diversity of the temporal graph of lifetime one and containing $G$ as unique snapshot. Hence, in the following, given a static graph $G$, we also use the notation $\tndg[G]$ to refer to the temporal neighbourhood diversity graph of the temporal graph containing $G$ as unique snapshot.

%
Let $X$ be a set of vertices and $Y$ be a set of classes. We say that $X$ and $Y$ are \emph{compatible} if each class $V_i$ belongs to $Y$ if and only if $V_i$ intersects $X$, that is, $\forall V_i: (V_i \cap X \neq \emptyset \Leftrightarrow V_i \in Y)$.
% we have $X \cap V_i \neq \emptyset \Leftrightarrow V_i \in Y$
Notice that there is exactly one subset of classes that is compatible with a set of vertices $X$ whereas several subsets of vertices of $G$ can be compatible with a set of classes $Y$.

We now introduce the concept of a neighbourhood diversity locally decidable problem -- we do this first in the static setting in order to build into the temporal setting. 
Intuitively, these are problems for which, given a set of classes $Y$, we can determine the minimum and maximum number of vertices to select in each class such that there exists at least one set of vertices with the desired property compatible with $Y$, if such a set exists. The formal definition is as follows:

\begin{definition}[Neighbourhood diversity locally decidable]
  \label{def:ndld}
  A static vertex-selection problem $\Pi_S$ with property $\Pi$ is $f(|G|)$-\emph{neighbourhood diversity locally decidable} if,
  for any static graph $G$ with $n$ vertices there are three computable functions 
  \begin{enumerate}
      \item $check_{\Pi}:\mathcal{P}(V(\ndg[G]))\to \{\textsc{True},\textsc{False}\}$,
      \item $low_{\Pi }: \mathcal{P}(V(\ndg[G])) \times V(\ndg[G]) \to \mathbb{N}$, and
      \item  $up_{\Pi} : \mathcal{P}(V(\ndg[G])) \times  V(\ndg[G]) \to \mathbb{N}$
  \end{enumerate} 
  with time complexity $\mathcal{O}(f(|G|))$ such that, for
    every subset of classes $Y$ of $\ndg[G]$, the following two conditions hold:
  \begin{enumerate}
    \item[(a)] $check_{\Pi}(Y)$ determines if there is a set respecting $\Pi$ in $G$ that is compatible with $Y$,
    \item[(b)] if there is such a set, for all subsets $X \subseteq V(G)$, $X$ is a set respecting $\Pi$ and is compatible with $Y$ if and only if, for all $V_i \in V(\ndg[G])$, $low_{\Pi}(Y,V_i) \leq |V_i \cap X| \leq up_{\Pi}(Y,V_i)$.
  \end{enumerate}
\end{definition}

  In other words, $low_{\Pi}$ and $up_{\Pi}$ are necessary and sufficient lower and upper bounds for the number of selected vertices in each class to respect the desired property in the graph of low neighbourhood diversity.

Notice that it is easy to compute a solution of minimum (respectively maximum) size that is compatible with a subset of classes $Y$ (if such a solution exists), by arbitrarily selecting exactly $low_{\Pi}(Y,V_i)$ (resp. $up_{\Pi}(Y,V_i)$) vertices inside each class $V_i$.
Hence, $\Pi_S$ is solvable in $\mathcal{O}(2^k \cdot f(|G|))$, where $k$ is the neighbourhood diversity of the graph. Indeed, it suffices to enumerate every subset of classes and keep the best solution. It follows that if $check_{\Pi}, low_{\Pi}$ and $up_{\Pi}$ are polynomial time functions, $\Pi_S$ is in FPT when parameterised by neighbourhood diversity.

%\subsubsection{Example problems that are neighbourhood diversity locally decidable}

We show {\sc Dominating Set} and {\sc Independent Set} are $f(n+m)$-neighbourhood diversity locally decidable.

\begin{lemmarep}
  \label{lemma:ds is ndld} {\sc Dominating set} is $f(|G|)$-neighbourhood diversity locally decidable where $f(|G|) \in O(n+m)$.
\end{lemmarep}

\begin{proof}
  Let $G$ be a static graph and let $Y \subseteq V(\ndg[G])$ be a subset of classes. We show that the following two conditions hold.
  \begin{enumerate}
    \item[(a)] there is a dominating set compatible with $Y$ if and only if $Y$ is a dominating set in $\ndg[G]$, and
    \item[(b)] if $Y$ is a dominating set in $\ndg[G]$, then for each $V_i \in \ndg[G]$ the lower and upper bounds functions $low : V(\ndg[G])  \to \mathbb{N}$ and $up : V(\ndg[G]) \to \mathbb{N}$ are given by
    \begin{enumerate}
      \item[(i)] $low(V_i) = up(V_i) =0$, if $V_i\not\in Y$,
      \item[(ii)] $low(V_i) = 1$ and $up(V_i) = |V_i|$, if $V_i\in Y$ and $V_i$ is a clique-class or has a neighbour in $Y$ and,
      \item[(iii)] $low(V_i) = up(V_i) =|V_i|$, otherwise.
    \end{enumerate}
  \end{enumerate}
  
  First, we show that there is a dominating set in $G$ compatible with $Y$ if and only if $Y$ is a dominating set in $\ndg[G]$.
  Let $Y$ be a dominating set in $\ndg[G]$, we show that the subset of vertices $X = \cup_{V_i \in Y} V_i$ is a dominating set of $G$. For any vertex $v \not \in X$ of $G$, the class $V_i \in V(\ndg[G])$ to which $v$ belongs has a neighbour $V_j$ in $\ndg[G]$ that belongs to $Y$ since otherwise $Y$ would not be a dominating set of $\ndg[G]$. Then, any vertex of $V_j$ belongs to $X$ and is a neighbour of $v$ in $G$ and so, $v$ is dominated. Hence, $X$ is a dominating set of $G$.

  Now, let $X$ be a dominating set of $G$, we show that the subset of classes $Y$ compatible with $X$ is a dominating set of $\ndg[G]$. For any class $V_i \not \in Y$ and any vertex $v \in V_i$, $v$ is dominated by a vertex $u$ that belongs to a class $V_j \neq V_i$. Then, $V_j \in Y$ and $V_j$ is adjacent to $V_i$ in $\ndg[G]$ and so, $V_i$ is dominated. Hence, any class that does not belong to $Y$ has a neighbour in $Y$ and so, $Y$ is a dominating set in $\ndg[G]$.

Further, suppose that $Y$ is a dominating set in $\ndg[G]$, % we show that the depicted bounds are necessary and sufficient.
and let $X$ be a dominating set of $G$ that is compatible with $Y$, we show that for all classes $V_i$ of $G$, we have $low_\Pi(V_i) \leq |X \cap V_i| \leq up_\Pi(V_i)$.
  Let $V_i$ be a class of $G$.
  If $V_i\not\in Y$, then we have $|X \cap V_i| = 0$ since otherwise $X$ would not be compatible with $Y$. 
  If $V_i\in Y$, since $X$ is compatible with $Y$, $X$ should contain at least one vertex in $V_i$. Moreover, since $X$ cannot contain more that $|V_i|$ vertices in $V_i$, we have $1 \leq |V_i \cap X| \leq |V_i|$.
  If $V_i$ is not a clique-class or has no neighbour in $Y$, then any vertex $v \in V_i$ is isolated in the subgraph induced by $X \cap \cup_{V_j \in Y} V_j$. Hence, $X$ contains necessarily every vertex of $V_i$ and so, $|X \cap V_i| = |V_i|$.

  Finally, we show that if $Y$ is a dominating set in $\ndg[G]$, then any set of vertices $X$ such that for all classes $V_i$, we have $low_\Pi(V_i) \leq |X \cap V_i| \leq up_\Pi(V_i)$,  is a dominating set in $G$ compatible with $Y$. First, since for any class $V_i$ we have $|V_i \cap X| = 0$ if $V_i \not\in Y$ and $1 \leq |V_i \cap X|$ otherwise, $X$ is compatible with $Y$. It remains to show that $X$ is also a dominating set.
  Let $v \not\in X$ be a vertex of $G$ that belongs to a class $V_i$. If $V_i \not\in Y$, then since $Y$ is a dominating set of $\ndg[G]$,  there is a class $V_j \in Y$ that is adjacent to $V_i$ in $\ndg[G]$. Thus, $V_j$ contains a vertex $u \in X$ and $u$ is adjacent to $v$ in $G$ and dominates $v$.
  If $V_i \in Y$ and is a clique-class, then there is a vertex $u \in V_i \cap X$ and $u$ is adjacent to $v$ in $G$. If $V_i$ is an independent-class, then since $v$ does not belong to $X$, there is necessarily another class $V_j \in Y$ that is adjacent to $V_i$ in $\ndg[G]$, since otherwise, every vertex of $V_i$ would belong to $X$. Thus, $V_j$ contains a vertex $u \in X$ and $u$ is adjacent to $v$ in $G$ and dominates $v$. Hence, in any case, $v$ is dominated by another vertex and thus, $X$ is a dominating set of $G$ that is compatible with $Y$.

  Note that we can compute $check(Y)$ in $O(m)$ and $low(V_i)$ and $up(V_i)$ in $O(n)$ for all $Y$ and $V_i$. This gives us the desired result.
\end{proof}

\begin{lemmarep}
  \label{lemma:is is ndld} {\sc Independent Set} is $f(|G|)$-neighbourhood diversity locally decidable where $f(|G|) \in O(n+m)$. 
\end{lemmarep}

\begin{proof}
    Let $G$ be a static graph and let $Y \subseteq V(\ndg[G])$ be a subset of classes. We show the following two conditions.
  \begin{enumerate}
    \item[(a)] there is an independent set compatible with $Y$ if and only if $Y$ is an independent set in $\ndg[G]$, and
    \item[(b)] if $Y$ is a independent set in $\ndg[G]$, then the lower and upper bounds functions $low : V(\ndg[G])  \to \mathbb{N}$ and $up : V(\ndg[G]) \to \mathbb{N}$ are given by
    \begin{enumerate}
      \item[(i)] $low(V_i) = up(V_i) =0$, if $V_i\not\in Y$,
      \item[(ii)] $low(V_i) = up(V_i) = 1$, if $V_i\in Y$ and $V_i$ is a clique-class and,
      \item[(iii)] $low(V_i)=1$ and $up(V_i) =|V_i|$, otherwise.
    \end{enumerate}

  \end{enumerate}
  
  First, we show that there is an independent set in $G$ compatible with a subset of classes $Y$ if and only if $Y$ is an independent set in $\ndg[G]$.
  Let $Y$ be an independent set in $\ndg[G]$, let $X \subseteq V(G)$ be any subset of vertices constructed by arbitrarily choosing one vertex in each $V_i \in Y$.
  We show that $X$ is an independent set of $G$. Let $u$ be a vertex of $X$ that belongs to some class $V_i$. Per construction of $X$, we have $V_i \in Y$. Let $v$ be a neighbour of $u$. If $v \in V_i$, then $v$ does not belong to $X$ since $X$ contains at most one vertex per class in $Y$. If $v$ belongs to another class $V_j$, then $V_i$ and $V_j$ are adjacent in \ndg[G] and since $Y$ is an independent set, $V_j \not \in Y$ and so, by construction of $X$, $v \not\in X$. So, $X$ is an independent set.
  Now, let $X$ be an independent set of $G$, we show that the subset of classes $Y$ compatible with $X$ is an independent set of $\ndg[G]$. Suppose there exist two adjacent classes $V_i$ and $V_j$ that belong to $Y$. Then, $V_i$ contains a vertex $u \in X$ and $V_j$ contains a vertex $v \in X$. Since $u$ and $v$ are adjacent, it contradicts $X$ being an independent set. Hence, $Y$ does not contain two adjacent classes and so, $Y$ is an independent set in \ndg[G].

Further, suppose that $Y$ is an independent set in $\ndg[G]$, % we show that the depicted bounds are necessary and sufficient.
and let $X$ be an independent set of $G$ that is compatible with $Y$, we show that for all classes $V_i$ of $G$, we have $low_\Pi(V_i) \leq |X \cap V_i| \leq up_\Pi(V_i)$.
  Let $V_i$ be a class of $G$.
  If $V_i\not\in Y$, then we have $|X \cap V_i| = 0$ since otherwise $X$ would not be compatible with $Y$. 
  If $V_i\in Y$, since $X$ is compatible with $Y$, $X$ should contain at least one vertex in $V_i$. Moreover, since $X$ cannot contain more that $|V_i|$ vertices in $V_i$, we have $1 \leq |V_i \cap X| \leq |V_i|$.
  If $V_i$ is a clique-class, then $X$ cannot contain more two vertices in $V_i$ since they would be adjacent.
  Hence, $|V_i \cap X|=1$ .

  Finally, we show that if $Y$ is an independent set in $\ndg[G]$, then any set of vertices $X$ such that for all classes $V_i$, we have $low_\Pi(V_i) \leq |X \cap V_i| \leq up_\Pi(V_i)$, is an independent set in $G$ compatible with $Y$. First, since for any class $V_i$ we have $|V_i \cap X| = 0$ if $V_i \not\in Y$ and $1 \leq |V_i \cap X|$ otherwise, $X$ is compatible with $Y$. It remains to show that $X$ is also an independent set.
  Suppose there exist two adjacent vertices $u$ and $v$ that both belong to $X$. The two vertices cannot belong to different classes $V_i$ and $V_j$ since otherwise both $V_i$ and $V_j$ would belong to $Y$, contradicting $Y$ being an independent set in \ndg[G]. Thus, $u$ and $v$ belong to the same class $V_i$. Since, $u$ and $v$ are adjacent, $V_i$ is not an independent class. But then, we have $|X \cap V_i| > up(V_i)$ which is a contradiction. Hence, $X$ does not contain two adjacent vertices and therefore $X$ is an independent set. 

  Note that we can compute $check(Y)$ in $O(m)$ and $low(V_i)$ and $up(V_i)$ in $O(n)$ for all $Y$ and $V_i$. This gives us the desired result.
\end{proof}

\subsubsection{In temporal graphs}
We now extend the concept of neighbourhood diversity locally decidable problems to a temporal setting. We use temporal neighbourhood diversity introduced by Enright et al.~\cite{abs-2404-19453}.

\begin{definition}[Temporal Neighbourhood Diversity~\cite{abs-2404-19453}]
The \emph{temporal neighbourhood diversity} of a temporal graph $\mathcal{G}=(G,\lambda)$ is the minimum value $k$ such that the vertices of $G$ can be partitioned into $k$ classes $V_1,\dots,V_k$ such that 
  for each pair of vertices $u$ and $v$ in a class $V_i$, we have $N_t(u) \setminus \{v\} = N_t(v) \setminus \{u\}$ at each time-step $t \in [1,\tau]$.  We call $V_1,\dots,V_k$ a \emph{temporal neighbourhood diversity partition} of $\mathcal{G}$.
\end{definition}

\noindent As for the static parameter, we can use a graph to represent the partition.

\begin{definition}[Temporal neighbourhood diversity graph]
 Let $\mathcal{G}=(G,\lambda)$ be a temporal graph with temporal neighbourhood partition $V_1,\dots,V_k$. The \emph{temporal neighbourhood diversity graph} of $\mathcal{G}$, denoted $\tndg$ is the temporal graph obtained by merging every class $V_i$ into a single vertex. Formally, we have $V(\tndg) = \{V_1,\dots,V_k\}$, $E(\tndg) = \{V_iV_j \mid \forall v_i \in V_i, \forall v_j\in V_j, v_iv_j \in E(G)\}$ and $\lambda(V_iV_j) = \lambda(u_iu_j)$ for any vertices $u_i \in V_i$ and $u_j \in V_j$.
\end{definition}

\begin{figure}
  \centering
  \scalebox{0.45}{
    \begin{tikzpicture}
      \newcounter{vertex}
      \setcounter{vertex}{1}

      \newcounter{tmp}
      \setcounter{tmp}{1}
      
    \foreach[count=\l from 1] \x/\T in {3/{C,I,C},4/{I,I,C},6/{I,I,I},6/{C,I,I},5/{C,C,I},2/{I,I,C}}{
      
      \setcounter{tmp}{\thevertex}

      \foreach[count=\X from 1] \i in \T {
        \setcounter{vertex}{\thetmp}
        \node[draw,circle,minimum width=2.1cm,label=\l*60+30:{\LARGE $V_\l$}] (\l\X) at ($(\X*9,0)+(\l*60:2.5)$) {};
        \pgfmathsetmacro{\a}{360/\x}% 
        \foreach \v in {1,...,\x}{
          \node[smallvertex,label=\v*\a:{\letter{\thevertex}}] (\v) at ($(\X*9,0)+(\l*60:2.5)+(\v*\a:0.5)$) {};
          \stepcounter{vertex}
        }
        \ifthenelse{\equal{\i}{C}}{
          \foreach \a in {1,...,\x}{
            \foreach \b in {1,...,\a} {
              \draw (\a)--(\b);
            }
          }
        }{}
      }
    }
    \foreach \a/\b in {21/31,41/51,41/11,41/61,51/61,11/61,11/51,22/12,62/12,22/62,62/52,22/42,42/52,23/33,53/63,53/13,53/43}
    \draw (\a)--(\b);
    \foreach \L in {1,...,3}
    \node at (10*\L,-5) {\LARGE $G_\L$};
  \end{tikzpicture}
  }
  \caption{\label{fig:tmp neighbourhood diversity} Example of temporal neighbourhood diversity graph \tndg of a temporal graph $\mathcal{G}=(G,\lambda)$. An edge between two sets $V_i$ and $V_j$ indicates that each vertex of $V_i$ is adjacent to all vertices of $V_j$.} 
\end{figure}
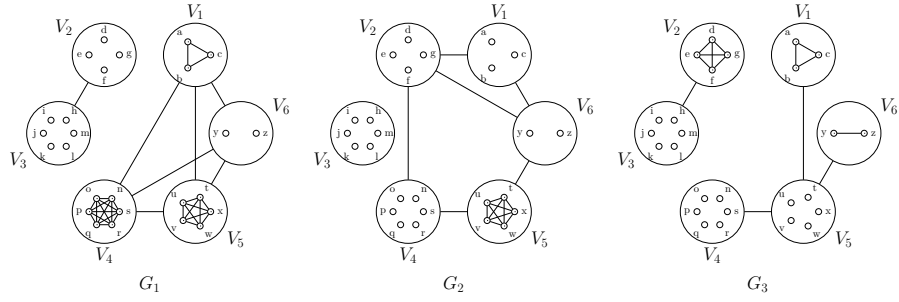
 
An example of a temporal neighbourhood diversity graph is depicted in \Cref{fig:tmp neighbourhood diversity}.
Let $\Pi_T$ be a temporally satisfying reconfiguration problem. We call $\Pi_T$ \emph{$f(|G|)$-temporal neighbourhood diversity locally decidable} if the static version $\Pi_S$ of $\Pi_T$ is $f(|G|)$-neighbourhood diversity locally decidable. 
% In both of the problems we consider $f(n)=n$.
% , and we simply write ``temporal neighbourhood diversity locally decidable'' for convenience. 
For a temporal graph $\mathcal{G}=(G,\lambda)$, we denote by $low^t_{\Pi}$ and $up^t_{\Pi}$ the functions giving the lower and upper bounds on the number of selected vertices in a class $V_i$ needed to obtain a set respecting $\Pi$ in $G_t$, as in \Cref{def:ndld}.

\subsection{FPT algorithm with respect to temporal neighbourhood diversity and lifetime}
\label{subsec:algo}

We now formulate an FPT algorithm using temporal neighbourhood diversity and lifetime to solve an $f(|G|)$-temporal neighbourhood diversity locally decidable reconfiguration problem $\Pi_T$ in a temporal graph $\mathcal{G}=(G,\lambda)$. 
A \emph{candidate sequence} is a sequence $Y = (Y_1,\dots,Y_\tau)$ such that for each $t \in [\tau]$, $Y_t$ is a subset of classes of the temporal neighbourhood diversity partition. 
We say that $Y$ is \emph{valid} if $check_{\Pi}(Y_t)= \texttt{true}$ for all $t \in [\tau]$. Let $T=(T_1,\dots,T_\tau)$ be a reconfigurable sequence for $\mathcal{G}$. For the sake of simplicity, we say that the two sequences $Y$ and $T$ are compatible if for each $t \in [\tau]$, $T_t$ and $Y_t$ are compatible. 

The principle of the algorithm is to decide how many vertices must be selected in each class of the decomposition. 
The algorithm does this by iterating through each candidate sequence $Y$ and, if $Y$ is valid, computing an optimal reconfigurable sequence $T$ compatible with $Y$. Then, we return an optimal solution among all sets associated to valid candidate sequences.
Notice that there are at most $2^{tnd \cdot\tau}$ candidate sequences to consider, where $tnd$ is the temporal neighbourhood diversity of the input graph. Hence, we can expect the algorithm to be efficient if the temporal graph has small temporal neighbourhood diversity and short lifetime.
Observe that a natural approach would be to enumerate feasible class subsets at each time step and apply local transitions between them using \Cref{lemma:enumeration}. However, this fails in general: the number of tokens needed in future snapshots may exceed the capacity of some classes, creating a bottleneck that prevents a valid reconfiguration, even if each local step appears feasible.
% This algorithm is depicted in \Cref{alg:fpt nd}.
If $\Pi_T$ is $f(|G|)$-temporal neighbourhood diversity locally decidable, we show in \Cref{subsec:reconfiguration} how to compute an optimal reconfigurable sequence compatible with a candidate sequence, if one exists, by solving an instance of the circulation problem. Combining the observation that there are at most $2^{tnd \cdot\tau}$ candidate sequences to consider and the result from Subsection~\ref{subsec:reconfiguration} gives us the following result, which we restate at the end of that subsection.

% \begin{algorithm}
%   \SetAlgoLined
%   \KwData{A temporal graph $\mathcal{G}$ and a reconfiguration problem $\Pi$}
%   \KwResult{The score of the optimal solution for $\Pi$ in $\mathcal{G}$}
%   \ForEach{candidate sequence $Y = (Y_1,\dots,T_\tau)$}{
%     \If{$check(Y)$}{
%       \tcc{compute an optimal sequence compatible with $Y$}
%       $T \gets compute\_sequence(\mathcal{G},Y)$\;
%       \lIf{$\Pi$ is a minimisation problem}{$score \gets \min(score,|T|)$}
%       \lElse{$score \gets \max(score,|T|)$}
%     }
%   }
%     \Return score\;
%   \caption{}
%   \label{alg:fpt nd}
% \end{algorithm}

\begin{lemma}
    Let $\mathcal{G}=(G,\lambda)$ be a temporal graph and let $\Pi_T$ be an $f(|G|)$-temporal neighbourhood diversity locally decidable reconfiguration problem. 
 $\Pi_T$ is solvable in $\mathcal{O}(2^{(tnd\cdot \tau)} \cdot (\tau\cdot tnd \cdot f(|G|) + \tau^{\nicefrac{11}{2}} \cdot tnd^8))$ time where $tnd$ is the temporal neighbourhood diversity of $\mathcal{G}$.
\end{lemma}

%%% Local Variables:
%%% mode: LaTeX
%%% TeX-master: "main"
%%% End:

\subsection{Computing an optimal reconfigurable sequence compatible with a candidate sequence}
\label{subsec:reconfiguration}

In this subsection, we present an efficient method for computing an optimal reconfigurable sequence $T$ compatible with a particular valid sequence $Y$. 
 Note that it is inefficient to compute $T$ by exhaustively generating all possible sequences in $\mathcal{G}$ and selecting the best one compatible with $Y$, as this would have time complexity of $\mathcal{O}(2^{\tau\cdot|V(G)|})$ for each candidate sequence making it impractical.
%
 % Given a valid candidate sequence $Y$, we describe how to compute an optimal reconfigurable sequence $T = (T_1,\dots,T_\tau)$ for $\Pi$ compatible with $Y$ in $\mathcal{O}(\tau\cdot tnd \cdot f(n) + \tau^{\nicefrac{11}{2}} \cdot tnd^8)$ time. % It corresponds to the function compute\_sequence of \Cref{alg:fpt nd}.

%

Our algorithm uses an instance of the circulation problem as a subroutine.
The circulation problem is a variation of the network flow problem on digraphs for which there are upper and lower bounds on the capacities of the arcs. 
%\begin{toappendix}

%\subsection{Computing an optimal reconfigurable sequence compatible with a candidate sequence}
    
Formally, a \emph{circulation graph} $\overrightarrow{F}=(F,c,l,u)$ consists of a directed graph $F$, along with a cost function $c : A(F) \to \mathbb{N}$ on the arcs and two functions on the arcs $l : A(F) \to \mathbb{N}$ and $u : A(F) \to \mathbb{N}$, representing the lower and upper bounds of the flow capacities. The digraph $F$ also has two designated vertices $src$ and $sink$ representing the source and the sink vertices of the flow, respectively.

In the circulation problem, we want to find a \emph{circulation flow} between $src$ and $sink$, which is a function $g : A(F) \to \mathbb{N}$ that respects two constraints:
\begin{enumerate}
\item[(a)] \emph{Flow preservation}: the flow entering in any vertex $v \in V(F) \setminus \{src,sink\}$ is equal to the flow leaving it, that is, $\sum\limits_{u \in N^+(v)} g(u,v) = \sum\limits_{u \in N^-(v)} g(v,u)$.
  \item[(b)] \emph{Capacity:} for all arcs $(u,v) \in A(F)$, the amount of flow over $(u,v)$ must respect the capacity of $(u,v)$, that is, $l(u,v) \leq g(u,v) \leq u(u,v)$.
\end{enumerate}
%
%\end{toappendix}
Formally, the circulation problem is defined as follows.
\prob{Circulation problem}{
A circulation digraph $\overrightarrow{F}$ with cost $c(a)$, upper capacity $u(a)$ and lower capacity $l(a)$ on each arc $a$, a source vertex $src$ and a sink vertex $sink$.
}{
  Find a circulation flow $g$ that minimises $\sum\limits_{a \in A(F)}g(a)\cdot c(a)$.
}
The circulation problem is known to be solvable in $\mathcal{O}(|A(F)|^{\nicefrac{5}{2}}\cdot |V(F)|^3)$~\cite{Tardos85}. Notice that the circulation problem can easily be transformed into a maximisation problem by modifying all positive costs into negative costs.

\begin{toappendix}
We now show that the problem of finding an optimal reconfigurable sequence compatible with $Y$ is reducible to the circulation problem.
Intuitively, we encode a candidate sequence for a temporally satisfying reconfiguration problem as an instance of the circulation problem in which each timestep is encoded as two layers of the digraph in the circulation problem and the arcs and capacities model the reconfigurability constraints between timesteps as well as the requirement that a set of vertices meets the required property at each time.

% \begin{construction}
%   \TD{constructive version\\}
%   \label{const:flow}
%   Let $\mathcal{G}=(G,\lambda)$ be a temporal graph, let $Y=(Y_1,\dots,Y_\tau)$ be a valid candidate sequence.
%   We construct the following circulation graph $\overrightarrow{F}=(F,c,l,u)$.
%   \begin{itemize}
%     \item For each $Y_t$ and each set $V_i \in Y_t$, introduce two vertices $x^t_i, y^t_i$ along with the arc $(x^t_i,y^t_i)$ and set $c(x^t_i,y^t_i)=0$, $l(x^t_i,y^t_i)=low_\pi(Y_t,V_i)$ and $u(x^t_i,y^t_i)=up_\pi(Y_t,V_i)$.
%     \item Introduce a source vertex $sou$ and for each set $V_i \in Y_1$, introduce the arc $(sou,x^1_i)$ and set $c(sou,x^1_i)=1, l(sou,x^1_i)=0$ and $u(sou,x^1_i)=+\infty$.
%     \item Introduce a target vertex $tar$ and for each set $V_i \in Y_\tau$, introduce the arc $(x^\tau_i,tar)$ and set $c(x^\tau_i,tar)=l(x^\tau_i,tar)=0$ and $u(x^\tau_i,tar)=+\infty$.
%     \item For each $t\in[1,\tau-1]$, and each pair $V_i \in Y_t, V_j \in Y_{t+1}$ such that either $V_i=V_j$ or $V_i$ and $V_j$ are adjacent in $\tndg$ at time step $t$, introduce an arc $(y^t_i,x^{t+1}_j)$ and set $c(y^t_i,x^{t+1}_j)=0, l(y^t_i,x^{t+1}_j)=0$ and $u(y^t_i,x^{t+1}_j)=+\infty$.  
%   \end{itemize}
% \end{construction}

\begin{construction}
  \label{const:flow}
  Let $\mathcal{G}=(G,\lambda)$ be a temporal graph, let $Y=(Y_1,\dots,Y_\tau)$ be a valid candidate sequence.  
  The reconfiguring circulation graph $\overrightarrow{RF}(\mathcal{G},Y)=(F,c,l,u)$ is the circulation graph with vertex set $V(F) = \{src,sink\} \cup \{x_i^t,y_i^t \mid V_i \in V(\tndg), t \in [1,\tau]\}$ and arc set $E(F) = S \cup \{C_t \mid t \in [1,\tau]\} \cup \{R_t \mid t \in [1,\tau-1]\}  \cup T$ where the arc sets $S,C_t ,R_t$ and $T$ are defined as follows: 
  \begin{itemize}
    \item $S= \{ (src,x_i^1) \mid V_i \in Y_1\}$ with $c(a)=1$ and $l(a)=0$ and $u(a)=+\infty$ for each a $\in S$,
    \item $C_t= \{(x^t_i,y^t_i) \mid V_i \in Y_t\}$ with $c(x^t_i,y^t_i) = 0$, $l(x^t_i,y^t_i) = low^t_\Pi(V_i)$ and  $u(x^t_i,y^t_i) = up^t_\Pi(V_i)$ for each $(x^t_i,y^t_i)\in C_t$,
    \item $R_t= \{(y^t_i,x^{t+1}_i) \mid V_i \in Y_t \cap Y_{t+1}\} \cup \{(y^t_i,x^{t+1}_j) \mid V_i \in Y_t,V_j \in Y_{t+1}, V_i$ is adjacent to $V_j$ in $\tndg$ at time $t\}$ with $c(a)=l(a)= 0$ and $u(a)=+\infty$ for each $a\in R_t$,
    \item $T = \{(y_i^\tau,sink) \mid V_i \in V(\tndg)\}$ with $c(a)=l(a)=0$ and $u(a)=+\infty$ for each $a\in T$.
  \end{itemize}
  \end{construction}

An example construction can be found in Figure~\ref{fig:flow}. The idea of the reduction is to represent each token by a unit value of the flow.
  At each time-step $t$, we can decompose the reconfiguring circulation graph into two parts: the checking part $C_t$ and the reconfiguring part $R_t$ (which only exists if $t \neq \tau$). In the checking part, each class $V_i$ is represented by an arc $(x^t_i,y^t_i) \in C_t$ and the value of the flow passing by this arc corresponds to the number of tokens inside $V_i$ at time-step $t$. Since the lower and upper bounds of the arc are the same as those described in \Cref{def:ndld}, we ensure that at each step, the set of selected vertices respects the property. In the 
reconfiguring part, an arc $(y^t_i,x^t_j) \in R_t$ indicates that the tokens contained in the class $V_i$ can move to the class $V_j$ during time-step $t$. An arc $(y^t_i,x^t_i) \in R_t$ indicates that the tokens in the class $V_i$ can stay in the same vertex during time-step $t$.

\begin{figure}[ht]
  \centering
  \scalebox{0.75}{
    \begin{tikzpicture}[
tap/.style args = {#1}{decoration={raise=5,
                                      text along path,
                                      text align={align=center},
                                      text={#1}
                                      },
              postaction={decorate},
              },
                   ]
   
     \foreach \x/\y/\l/\c/\u/\t in {0/0/1/1/3/1,0/-2/2/1/4/1,0/-4/3/1/6/1,
       4/1/6/1/3/2,4/-2/2/1/4/2,4/-4/3/6/6/2,
       8/-2/2/1/4/3,8/-4/3/1/6/3,8/1/6/1/3/3,8/-1/5/1/5/3}
     {
       \draw[->,>=stealth] (\x,\y) node[smallvertex,label=90:{$x_\l^\t$}] (x\l\t) {} ->
       node[midway,above] {0/\c/\u}
       ++(1.9,0) node[smallvertex,label=90:{$y_\l^\t$}] (y\l\t) at ++(0.1,0) {}
       ;
     }
     \foreach \a/\b in {21/22,22/23,31/32,32/33,21/32,31/22,11/62,62/63,22/63,22/53,62/53}
     \draw[->] (y\a)->(x\b);

     \node[smallvertex,label=90:{$src$}] (s) at (-3,-2) {};
     \foreach \a in {11,21,31}
     \draw[->,tap={1/0/${+\infty}${}}] (s) -- node[midway,above] {} (x\a);

     \node[smallvertex,label=-90:{$sink$}] (t) at (13,-2) {};
     \foreach \a in {63,53,23,33}
     \draw[->] (y\a)--(t);
   \end{tikzpicture}
   }
   \caption{\label{fig:flow} Example of flow graph produced by \Cref{const:flow}. The input temporal graph is the one depicted in \Cref{fig:tmp neighbourhood diversity}. We want to compute a minimum dominating set compatible with the candidate sequence $Y = (Y_1 = \{V_1,V_2,V_3\}, Y_2= \{V_2,V_3,V_6\},Y_3 = \{V_2,V_3,V_5,V_6\})$.}
 \end{figure}
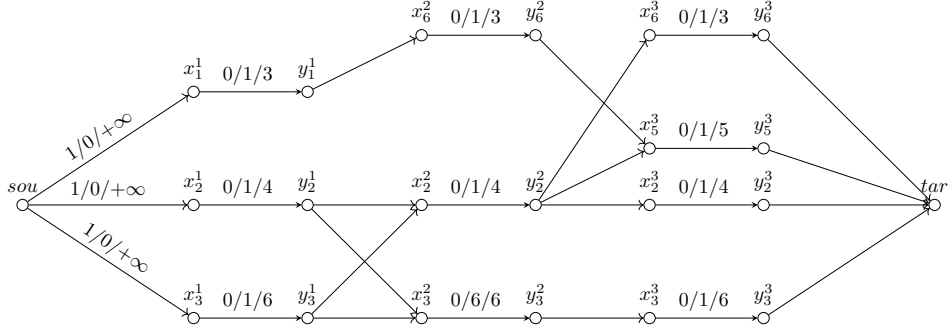

\begin{lemmarep}
  \label{lemma:flow reconfiguration} Let $\mathcal{G}$ be a temporal graph with temporal neighbourhood graph \tndg and let $Y=(Y_1,\dots,Y_\tau)$ be a valid candidate sequence for some $f(|G|)$- temporal neighbourhood diversity locally decidable reconfiguration problem $\Pi_T$. Let $\overrightarrow{RF}(\mathcal{G},Y)=(F,c,l,u)$ be the reconfiguring circulation graph as described in \Cref{const:flow}.
  There is a reconfigurable sequence $T$ compatible with $Y$ of size $\ell$ if and only if there is a circulation flow $g$ in $\overrightarrow{RF}(\mathcal{G},Y)$ with cost $\ell$.
  % $\mathcal{Y}$ is valid if there is a feasible flow $f$ in $\overrightarrow{F}$ and in that case, the minimum size of a reconfiguration sequence associated to $\mathcal{Y}$ is given by the minimum cost value for 

  % There is a sliding dominating set $(S,M,tok)$ in $\mathcal{G}$ with $\ell$ tokens if and only if there is a circulation flow $f$ with cost $\ell$ in $\overrightarrow{F}$.
\end{lemmarep}

\begin{proof}
  For any vertex $v$ of $\overrightarrow{RF}(\mathcal{G},Y)$, we denote $g^+(v)=\sum\limits_{u \in N^+(v)} g(u,v)$ and $g^-(v)= \sum\limits_{u \in N^-(v)} g(v,u)$, the flow entering in $v$ and leaving $v$, respectively.

  Let $T = (T_1,\dots,T_\tau)$ be a reconfigurable sequence of size $\ell$. For each time-step $t \in [1,\tau-1]$, let $b_t$ be a reconfiguration bijection between $T_t$ and $T_{t+1}$.  We construct a circulation flow $g$ between $src$ and $sink$ as follows.
  Let $V_i$ denote a class of \tndg.
  First, we set $g(src,x^1_i):= |T_1 \cap V_i|$.
  Then, for each time-step $t \in [1,\tau]$, we set $g(x^t_i,y^t_i) := |V_i \cap T_t|$ (i.e.\ the number of tokens contained in the class $V_i$ at time-step $t$).
  For each time-step $t \in [1,\tau-1]$, we set
    $g(y^t_i,x^{t+1}_j) := | \{ (u,v) \in b_t \mid u \in T_t \cap V_i, v \in T_{t+1} \cap V_j\} |$ (i.e.\ the number of tokens moving from the class $V_i$ to another class $V_j$ during time-step $t$ if $i \neq j$ or the number of tokens moving inside the class $V_i$ or staying in the same vertex in $V_i$ during time-step $t$ if $i = j$).
    %
    % $f(y^t_i,w^t_i) := | \{ uv \in M_t \mid u \in T_t \cap V_i, v \in T_{t+1} \cap V_i\} \cup T_t \cap T_{t+1} \cap V_i|$ (\textit{i.e.} the number of tokens moving inside the class $V_i$ or staying in the same vertex in $V_i$ during time-step $t$)
    %
    Finally, we set $g(y^\tau_i,sink):= |V_i \cap T_\tau|$. 
    We now show that $g$ is a circulation flow of cost $\ell$. First, notice that the only non-zero cost arcs are those leaving $src$. Hence, the cost of $g$ is $\sum\limits_{V_i\in Y_1} g(src,x^1_i) = \sum\limits_{V_i\in Y_1} |V_i \cap T_1| = \ell$.
    Further, we show that the capacity constraints are respected. For each arc $(x^t_i,y^t_i)$, since $\Pi_T$ is $f(|G|)$-temporal neighbourhood diversity locally decidable, we have $low^t_\Pi(Y_t,V_i) = l(x^t_i,y^t_i) \leq g(x^t_i,y^t_i) \leq up^t_\Pi(Y_t,V_i) = u(x^t_i,y^t_i)$. For any other arc $a$, we have $l(a)=0$ and $u(a)=+\infty$ and so, the capacity constraint is necessarily respected. Hence, the capacity constraints are respected for all arcs.
    It remains to show that $g$ respects the flow conservation constraint. 
    For any vertex $x^t_i$, we have $g^-(x^t_i) = g(x^t_i,y^t_i) = |T_t \cap V_i|$. If $t=1$, then $g^+(x^1_i)=g(src,x^1_i)= |T_1 \cap V^1_i|= g^+(x^t_i)$. Otherwise, by construction of $g$, $g^+(x^t_i)$ is equal to the sum of the number of tokens moving from another class $V_j$ to the class $V_i$ plus the number of tokens staying inside the class $V_i$ during the time-step $t$, that is, the number of tokens inside $V_i$ at time-step $t$. So, $g^-(x^t_i) = |T_t \cap V_i| = g^+(x^t_i)$. 
    For any vertex $y^t_i$, we have $g^+(y^t_i) = g(x^t_i,y^t_i) = |T_t \cap V^t_i|$. If $t = \tau$, then $g^-(y^t_i) = g(y^t_i,sink) = |T_\tau \cap V^\tau_i| = g^+(y^t_i)$. Otherwise, by construction of $g$, $g^-(y^t_i)$ is equal to the number of tokens moving from the class $V_i$ to another class $V_j$ plus the number of tokens staying inside the class $V_i$ during the time-step $t$, that is, the number of tokens inside $V_i$ at time-step $t$. So, $g^+(y^t_i) = |T_t \cap V_i| = g^-(y^t_i)$.

    We now show the reverse.
    Let $g$ be a circulation flow with cost $\ell$ in $\overrightarrow{RF}(\mathcal{G},Y)$.
    We construct a reconfigurable sequence $T = (T_1,\dots,T_\tau)$ compatible with $Y$ of size $\ell$ as follows. First, for each class $V_i \in Y_1$, we construct $T_1$, by selecting $g(x^1_i,y^1_i)$ vertices in the class $V_i$. Since $g$ respects the flow conservation constraint, we have $|T_1 \cap V_i| = g(x^1_i,y^1_i)$ for each class $V_i$. Moreover, since $low^1_\Pi(Y_1,V_i) = l(x^1_i,y^1_i) \leq g(x^1_i,y^1_i) < u(x^1_i,y^1_i) = up^1_\Pi(Y_1,V_i)$ and since $\Pi_T$ is $f(|G|)$-temporal neighbourhood diversity locally decidable, $T_1$ respects $\Pi$ in $G_1$.
    Now suppose that at time-step $t<\tau-1$, we also have $|T_t \cap V_i| = g(x^t_i,y^t_i)$ for each class $V_i$. We construct $T_{t+1}$ by moving $g(y^t_i,x^{t+1}_j)$ tokens from the class $V_i$ to the class $V_j$ for each pair $i,j$ such that $i\neq j$ and $(y^t_i,x^{t+1}_j)$ is an arc in $\overrightarrow{RF}(\mathcal{G},Y)$. Observe that inside $V_i$, exactly $g(y^t_i,x^{t+1}_i)$ tokens remain on the same vertex. The number of tokens leaving $V_i$ or staying inside $V_i$ is equal to $g^-(y^t_i)$ and since $g$ respects the flow conservation constraint, we have $g^+(y^t_i)= |T_t \cap V_i| = g^-(y^t_i)$. Thus, we have exactly the correct number of tokens in $V_i$ to make the moves. Moreover, we have  $g^+(x^{t+1}_j)= g(x^{t+1}_j,y^{t+1}_j)$ and so, we have $|T_{t+1} \cap V_j| = g(x^{t+1}_j,y^{t+1}_j)$. Since $low^{t+1}_\Pi(Y_{t+1},V_j) = l(x^{t+1}_j,y^{t+1}_j) \leq g(x^{t+1}_j,y^{t+1}_j) < u(x^{t+1}_j,y^{t+1}_j) = up^{t+1}_\Pi(Y_{t+1},V_i)$ and since $\Pi_T$ is $f(|G|)$-temporal neighbourhood diversity locally decidable, $T_{t+1}$ is a respects $\Pi$ in $G_{t+1}$. Hence, inductively, we have constructed a reconfigurable sequence $T = (T_1,\dots,T_\tau)$ compatible with $Y$.
\end{proof}

We can now conclude that it is possible to compute an optimal reconfigurable sequence compatible with a candidate sequence in polynomial time.
\end{toappendix}

This gives us the tools to produce the following Lemma:
 
\begin{lemmarep}
  \label{lemma:compute reconfigurable sequence} Let $\Pi_T$ be a $f(|G|)$-temporal neighbourhood diversity locally decidable problem, let $\mathcal{G}=(G,\lambda)$ be a temporal graph with lifetime $\tau$ and temporal neighbourhood diversity $tnd$ and let $Y$ be a valid candidate sequence. It is possible to compute an optimal reconfigurable sequence $T$ compatible with $Y$ in
$\mathcal{O}(\tau\cdot tnd \cdot f(|G|) + \tau^{\nicefrac{11}{2}} \cdot tnd^8)$ time.
\end{lemmarep}

\begin{proof}
  We first construct the reconfiguring circulation graph $\overrightarrow{RF}(\mathcal{G},Y)=(F,c,l,u)$, as depicted in \Cref{const:flow}. We have $|V(\overrightarrow{RF})|=\mathcal{O}(\tau\cdot \sum\limits_{Y_i \in Y}|Y_i|) = \mathcal{O}(\tau \cdot tnd)$ and $|A(\overrightarrow{RF})|= \mathcal{O}(\tau \cdot \sum\limits_{Y_i \in Y}|Y_i| + \sum\limits_{Y_i \in Y \setminus Y_\tau}(|Y_i|\cdot |Y_{i+1}|)) = \mathcal{O}(\tau \cdot tnd^2)$. Moreover, we call the two functions $low^t_\Pi$ and $up^t_\Pi$ exactly once per arc $(x^t_i,y^t_i)$ \textit{i.e.}, $\mathcal{O}(\tau\cdot tnd)$ calls in total. Hence,  the construction of $\overrightarrow{RF}$ can be done in $\mathcal{O}(\tau\cdot tnd \cdot f(|G|) + |A(\overrightarrow{RF})| + |V(\overrightarrow{RF})|)$.
  Then, we compute an optimal circulation flow $f$ in $\overrightarrow{RF}$ which can be done in $\mathcal{O}(|A(\overrightarrow{RF})|^{\nicefrac{5}{2}} \cdot |V(\overrightarrow{RF})|^3) = \mathcal{O}((\tau\cdot tnd^2)^{\nicefrac{5}{2}}\cdot (\tau\cdot tnd)^3) = \mathcal{O}(\tau^{\nicefrac{11}{2}}\cdot tnd^8)$~\cite{Tardos85}. We can then construct an optimal reconfigurable sequence $T$ compatible with $Y$, as described in the proof of \Cref{lemma:flow reconfiguration}. We obtain an overall complexity of $\mathcal{O}(\tau\cdot tnd \cdot f(|G|) + \tau^{\nicefrac{11}{2}} \cdot tnd^8)$.
\end{proof}

We now have the necessary tools for the overall algorithmic result:

\begin{lemmarep}   \label{lemma:fpt nd}
  Let $\mathcal{G}=(G,\lambda)$ be a temporal graph and let $\Pi_T$ be an $f(|G|)$-temporal neighbourhood diversity locally decidable reconfiguration problem. 
 $\Pi_T$ is solvable in $\mathcal{O}(2^{(tnd\cdot \tau)} \cdot (\tau\cdot tnd \cdot f(|G|) + \tau^{\nicefrac{11}{2}} \cdot tnd^8))$ time where $tnd$ is the temporal neighbourhood diversity of $\mathcal{G}=(G,\lambda)$.
\end{lemmarep}

\begin{proof}
  
  For each candidate sequence $Y = (Y_1,\dots,Y_\tau)$, we call the function $check_\Pi$ and we compute an optimal reconfigurable sequence compatible with it. Thus, each iteration can be done in $\mathcal{O}(\tau\cdot tnd \cdot f(|G|) + \tau^{\nicefrac{11}{2}} \cdot tnd^8)$. For each time-step $t$, there exists $2^{|V(\tndg|)} = 2^{(tnd)}$ possibilities to choose $Y_t$. Hence, there are $2^{tnd\cdot \tau}$ candidate sequences. We obtain an overall complexity of $\mathcal{O}(2^{tnd\cdot\tau} \cdot (f(|G|)))$.
\end{proof}

\begin{corollaryrep}
  Let $\mathcal{G}=(G,\lambda)$ be a temporal graph with temporal neighbourhood diversity $tnd$ and lifetime $\tau$.
  {\sc Temporal Dominating Set Reconfiguration} and {\sc Temporal Independent Set Reconfiguration} are solvable in $\mathcal{O}(2^{(tnd\cdot\tau)}\cdot(\tau^{\nicefrac{11}{2}} \cdot tnd^8))$ time in $\mathcal{G}$.
  
\end{corollaryrep}

\begin{proof}
By \Cref{lemma:ds is ndld} and~\Cref{lemma:is is ndld}, the complexity of the functions $check$ for $low_\Pi$ and $up_\Pi$ {\sc Dominating Set} and {\sc Independent Set} is $\mathcal{O}(|V(\tndg)|+|E(\tndg)|) = \mathcal{O}(tnd^2)$. Hence, by \Cref{lemma:compute reconfigurable sequence}, we can compute and check an optimal sequence in time $\mathcal{O}(\tau^{\nicefrac{11}{2}} \cdot tnd^8)$. Hence, by \Cref{lemma:fpt nd}, {\sc Temporal Dominating Set Reconfiguration} and {\sc Temporal Independent Set} are solvable in $\mathcal{O}(2^{(tnd\cdot\tau)}\cdot(\tau^{\nicefrac{11}{2}} \cdot tnd^8))$.
\end{proof}

\section{Fixed-parameter tractability with respect to lifetime and treewidth of footprint}

\label{sec:tw algo}
%In this section we state that if a static problem is FPT in the treewidth , then its temporally satisfying reconfiguration version is FPT in the treewidth of the footprint (the union of all time-steps) and the lifetime of the temporal graph.
Courcelle's celebrated theorem~\cite{Courcelle86a,Courcelle90} tells us that any static graph problem that can be encoded in monadic second-order logic (MSO) is in FPT with respect to the treewidth of the graph and the length of the MSO expression.  
We make use of this standard tool to argue that if a static problem $\Pi_S$ is definable in the monadic second-order logic, then its reconfiguration version $\Pi_T$ is FPT when parameterised by the treewidth of the footprint, the lifetime of the temporal graph and the length of the MSO expression describing $\Pi_s$. The definition of treewidth is given below (Definition~\ref{def:TD}). Roughly, it is a measure of how ``tree-like'' a static graph is.

\begin{toappendix}
We first introduce definitions specific to this section. Then, using Courcelle's theorem, we show that if the static version is expressible in MSO then the reconfiguration version can be parameterised by the lifetime and the treewidth.

%Finally, we show how to adapt a generic algorithm on a tree decomposition for the static version to the reconfiguration version.

% \subsection{Definitions}

\subsection{Treewidth and tree decompositions}

Tree decompositions are widely used to solve a large class of combinatorial problems efficiently by dynamic programming when the graph has low treewidth. 

\begin{definition}[Treewidth, tree decomposition \cite{BodlaenderK96,Kloks94}]\label{def:TD}
  Given a static graph $G$, a \emph{tree decomposition} of $G$ is a pair $(\mathcal{T},\mathcal{X})$
  where $\mathcal{T}$ is a tree and 
  $\mathcal{X}=\{B_i\mid i\in V(\mathcal{T})\}$ is a multiset of subsets of $V(G)$ (called ``bags'') such that
  \begin{enumerate}[(a)]
    \item for each $uv \in E(G)$, there is some $i$ with $uv \subseteq B_i$ and
    \item for each $v \in V(G)$, the bags $B_i$ containing $v$ form a connected subtree of $\mathcal{T}$.
    \end{enumerate}
    \vspace{0.1cm}
  The width of $(\mathcal{T},\mathcal{X})$ is $\max_{B_i \in \mathcal{X}} |B_i|-1$.
%   Further, $(\mathcal{T},\mathcal{X})$ is called \emph{nice} if
%   \begin{enumerate}[(i)]
%     \item $\mathcal{T}$ is rooted at bag $B_r$, with $B_r = \emptyset$ and each bag has at most two children.
%     \item Each bag $B_i$ of $\mathcal{T}$ has one of the four types:
%       \begin{itemize}
%       \item \textbf{Leaf bag:} $i$ has no children and $B_i = \emptyset$.
%         \item \textbf{Join bag:} $i$ has two children $j$ and $k$ and $B_i = B_j = B_k$.
%         \item \textbf{Introduce vertex $u$ bag:} $i$ has only one child $j$ and $B_j = B_i\setminus\{u\}$.
% %      \item \textbf{Introduce edge $uv$ bag:} $i$ has only one child $j$ and $B_j = B_i$.
%       \item \textbf{Forget $u$ bag:} $i$ has only one child $j$ and $B_j = B_i\cup\{u\}$.
%     \end{itemize}
%   \end{enumerate}
%   \vspace{0.1cm}
% 
  % For any bag $B_i$ of $\mathcal{T}$, we let $\mathcal{T}_i$ denote the subgraph of $G$ containing all vertices and edges that have been introduced ``below'' $B_i$
  % (that is, in a bag of the subtree of $T$ that is rooted at $i$).
\end{definition}

\noindent It is NP-complete to determine whether a graph has treewidth at most $k$~\cite{Arnborg87}. However, there exists a linear-time algorithm that, for any constant $k$, computes a tree-decomposition with treewidth at most $k$, if there is one~\cite{Bodlaender96}.

Monadic second-order logic (MSO logic) is a fragment of second-order logic where quantification is restricted to sets. The subscript MSO$_2$ is used to show we are allowed quantification over both vertex and edge sets. Importantly, if a graph property is expressible in MSO$_2$ logic, Courcelle's theorem states that there exists a fixed-parameter tractable algorithm in the treewidth of the graph and in the length of the MSO$_2$ expression.
In graphs, we are allowed to use the following variables and relations to express a property in MSO$_2$ logic:

\begin{itemize}
  \item standard boolean connectives: $\neg$ (negation), $\wedge$ (and), $\vee$ (or), $\Rightarrow$ (implication),
  \item standard quantifiers: $\exists$ (existential quantifier), $\forall$ (universal quantifier), which can be applied to any variable used to represent vertices, edges, sets of vertices or sets of edges of a graph. By convention, lower-case letters are used to represent vertices and edges, and upper-case letters are used to represent sets of vertices or edges,
  \item the binary equality relation $=$, the binary inclusion relation $\in$, the binary incidence relation $inc(e,v)$ which encodes that an edge $e$ is incident to a vertex $v$.
\end{itemize}

A \emph{free variable} is a variable not bound by quantifiers.  An MSO \emph{sentence} is an MSO formula with no free variables. 
Let $G$ be a graph, the notation $G\models \phi$ indicates that $G$ verifies the formula. 

% EMSO$_2$ is the extension of MSO$_2$ which allows for expression of a problem in MSO$_2$ and an evaluation function for sets which respect the desired property. This allows encoding of optimisation problems and decision problems with integer values in their input. We use this to eliminate the dependency of the length of our MSO$_2$ formulae on a factor of $k$ (size of set of selected vertices).

% \begin{definition}[\cite{langer2014practical}]
%   \label{def:evaluation relation}
%     An \emph{evaluation term} over variables $y_1\ldots,y_{\ell}$ is a function $t:\mathbb{R}^{\ell}\to \mathbb{R}$ built using arithmetic operators $-,+,\times$, the real numbers and a subset of $\{y_1,\ldots,y_{\ell}\}$. Let $\mathcal{T}_{y_1,\ldots,y_{\ell}}$ be the set of all evaluation terms over $y_1,\ldots,y_{\ell}$. An \emph{evaluation relation} over $y_1,\ldots,y_{\ell}$ is a propositional formula over the atomic formulas $\{t\leq 0, t=0 \,:t\in \mathcal{T}_{y_1,\ldots,y_{\ell}}\}$.
% \end{definition}
% Note that the above definition usually allows for functions which add weights to the vertices. This is not applicable in our setting, so it is omitted.

% Suppose we have a static decision vertex-selection problem that asks that the number of vertices selected is at most $k$ and the selected vertices satisfy a property expressible by the MSO$_2$ expression $\phi$. We can express this problem using $\phi$ and by making $y_1$ be the number of vertices selected, $y_2=k$, the evaluation term $t=y_1-y_2$, and the evaluation relation $\psi:=t\leq 0$.

\subsection{MSO formulation}
\end{toappendix}

\begin{theorem}[Courcelle's theorem~\cite{Courcelle86a,Courcelle90}]
  \label{theorem:courcelle} Let $G$ be a simple graph of treewidth $tw$ and $\phi$ be a fixed MSO$_2$ sentence. There exists an algorithm that tests if $G\models \phi$ and runs in $f(tw,|\phi|) \cdot |G|$ time, where $f$ is a computable function.
\end{theorem}

We now state an optimisation version of Courcelle's theorem, allowing us to circumvent dependency on the number of vertices selected. This uses a formula with free variables, rather than a sentence. This means that, for our purposes, we simply describe the property required of the set of selected vertices $X$ and optimise its cardinality instead of writing a sentence which quantifies over such a set.

\begin{toappendix}
% Here we use an extension of Courcelle's theorem~\cite{arnborg_easy_1991} shown by Arnborg, Lagergren and Seese which shows inclusion of problems expressible in EMSO$_2$ in FPT with respect to formula length and treewidth. Recall we make the assumption that the vertices are unweighted, and a vertex can be selected at most once in the problems we consider, we present a restricted version of the theorem statement given in~\cite{arnborg_easy_1991} for ease of reading.

\begin{theorem}[Adapted from \cite{arnborg_easy_1991}, as written in~\cite{zschoche2020complexity}]\label{thm:EMSO-opt}
    There exists an algorithm that, when given
    \begin{itemize}
        \item an MSO formula $\phi$ with free monadic variables $X_1,\ldots,X_r$,
        \item an affine function $\alpha(X_1,\ldots, X_r)$, and
        \item a graph $G$,
    \end{itemize}
    finds the minimum (maximum) of $\alpha(|X_1|,\ldots,|X_r|)$ over evaluations of $X_1,\ldots,X_r$ for which $\phi$ is satisfied on $G$. The algorithm requires $f(tw,|\phi|) \cdot |G|$ time, where $f$ is a computable function.
\end{theorem}

In order to apply the above theorem, we first convert the temporal graph $\mathcal{G}=(G,\lambda)$ into a static graph $H$ as defined in the following construction.

\begin{construction}
  \label{const:sliding}
  Given a temporal graph $\mathcal{G}=(G,\lambda)$ with bounded lifetime $\tau$ and vertex set $V(G) = \{u_1,\dots,u_n\}$, we consider the following static graph $H$ with vertex set $V(H) = \{ v^t_i \mid u_i \in V(G), t \in [1,\tau]\}$ and such that $H$ contains the following edges:% and edge set $E(H) = \bigcup\limits_{t \in [1,\tau]} E_t \bigcup\limits_{t \in [1,\tau-1]} E_{t,t+1}$ where $E_t = \{v^t_iv^t_j \mid u_iu_j \in G_t\}$ and $E_{t,t+1} = \{v^t_iv^{t+1}_j \mid u_iu_j \in G_t\} \cup \{v^t_iv^{t+1}_i \mid u_i \in V(G)\}$.
  \begin{itemize}
    \item for each $t \in [1,\tau]$, $H$ contains the set of edges $E_t = \{ v^t_iv^t_j \mid u_iu_j \in E(G_t)\}$ (\textit{i.e.} $H$ contains the disjoint union of each snapshot $G_t$), and
    \item for each $t \in [1,\tau-1]$, $H$ contains the edge set $E_{t,t+1} = \{v^t_iv^{t+1}_j \mid u_iu_j \in E(G_t)\} \cup \{v^t_iv^{t+1}_i \mid u_i \in V(G)\}$ (\textit{i.e.} there is an edge between $v^t_i$ and $v^{t+1}_j$ if it is possible to move a token from $u_i$ to $u_j$ at time-step $t$).  
  \end{itemize}
\end{construction}
Note that this construction is related to but is not the same as the time-expanded graph of a temporal graph (as used in, e.g.~\cite{fluschnik_as_2020}).

\begin{lemmarep}
Let $\mathcal{G}=(G,\lambda)$ be a temporal graph, and let $H$ be the graph described in \Cref{const:sliding}. The treewidth of $H$ is at most $\tau\cdot tw$ where $tw$ is the treewidth of the footprint $G$.
\end{lemmarep}
\begin{proof}
  Let $(T,\mathcal{X})$ be a tree decomposition of $G$ of width $tw$. We construct a tree decomposition $(T,\mathcal{Y})$ for $H$ as follows. For each bag $B_x \in \mathcal{X}$, we substitute each vertex $u_i \in B_x$ by the set of vertices $\{v^t_i \mid t \in [1,\tau]\}$. Clearly, for any vertex $v^t_i \in V(H)$, the bags in $\mathcal{Y}$ containing $v^t_i$ are connected in $T$ since the bags in $\mathcal{X}$ containing $u_i$ are connected in $T$.
For each edge $v^t_iv^t_j \in E(H)$ (respectively $v^t_iv^{t+1}_j$), let $B_x$ be a bag containing $u_iu_j$ in $\mathcal{X}$. After the substitution, $B_x$ contains the edge $v^t_iv^t_j$ (resp. $v^t_iv^{t+1}_j$). For each edge $v^t_iv^{t+1}_i$, any bag $B_x$ containing the vertex $u_i \in \mathcal{X}$ contains the edge $v^t_iv^{t+1}_i$. Hence, $(T,\mathcal{Y})$ is a tree decomposition of $H$ and since each vertex is replaced by $\tau$ vertices in each bag, the width of $(T,\mathcal{Y})$ is $\tau \cdot tw$.
\end{proof}

\noindent An example of a graph produced by \Cref{const:sliding} is depicted in \Cref{fig:to static}.

\begin{figure}[ht]
  \centering
  \scalebox{0.75}{
  \begin{tikzpicture}
    \foreach \Y in {1,2} {
      \foreach \X in {1,...,3} {
        \foreach[count=\l from 1] \x/\y/\a in {0/0/-90,1/-1/-90,0/2/90,1/1/-135,1/3/90}{
          \ifthenelse{\equal{\Y}{2}}{
            \node[smallvertex,label=90+90*\X:{$v^\X_\l$}] (\X\l) at (\X*2.5+\x+\Y*9,\y*0.75) {};
          }{
            \node[smallvertex,label=90+90*\X:{$u_\l$}] (\X\l) at (\X*2.5+\x+\Y*9,\y*0.75) {};
          }
        }

        \ifthenelse{\equal{\Y}{1}}{
          \node at (\X*2.5+0.5+\Y*9,-2) {$G_\X$};
        }
        {}
      }
      
      \foreach \a/\b/\L in {2/1/{2,3},1/3/{1,3},3/5/{1},3/4/{2}} {
        \foreach \X in \L {
          \draw (\X\a) -- (\X\b);
          \ifthenelse{\equal{\Y}{2}}{
          \ifthenelse{\equal{\X}{3}}{}{
            \pgfmathsetmacro{\B}{int(\X + 1)}
            \draw[draw=iris,dashed] (\X\a) -- (\B\b);
            \draw[draw=iris,dashed] (\B\a) -- (\X\b);
          }
          }{}
        }
      }
      \ifthenelse{\equal{\Y}{2}}{
        \foreach[count=\B from 2] \X in {1,2}{
          \foreach \a in {1,...,5}{
          \draw[draw=iris,dashed] (\X\a) -- (\B\a);
        }
        }
      }{}
      
    }
 % \foreach \a in {12,22,33}
 % \node[smallvertex,fill=gray] at (\a.center) {};
    
  \end{tikzpicture}
}
  \caption{\label{fig:to static} Example of a graph produced by \Cref{const:sliding}. The edges in $E_{t,t+1}$ sets are depicted in blue/dashed.}
\end{figure}
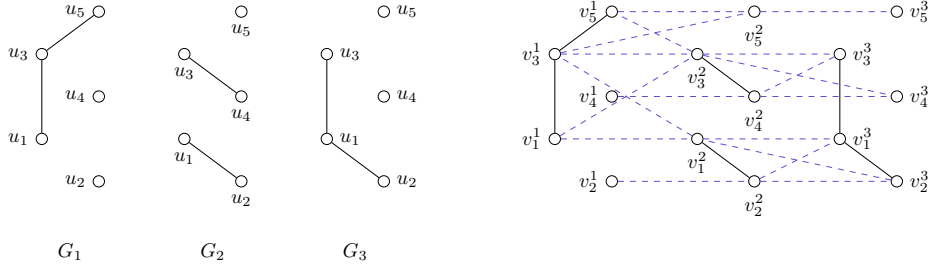

Let $\mathcal{G}$ be a temporal graph and let $H$ be the static graph produced by \Cref{const:sliding} with $\mathcal{G}$ given in input. For convenience, we denote by $V_t$ the set of vertices in $H$ with superscript $t$. Notice that finding a solution for $\Pi_T$ in $\mathcal{G}$ is equivalent to finding a vertex set $X$ in $H$ such that:
\begin{itemize}
  \item for each $t \in [\tau]$, $X \cap V_t$ is a set respecting $\Pi$ in $H[V_t]$, and
  \item for each $t \in [\tau-1]$, there is a perfect matching between $X \cap V_t$ and $X \cap V_{t+1}$.
  \end{itemize}
We show that if $\Pi$ can be expressed in MSO$_2$ logic, then we can also express the reconfiguration version $\Pi_T$ in MSO$_2$ logic, with an increase in formula size by a factor $\tau$.
By Courcelle's theorem, it follows that $\Pi_T$ is FPT in the length of the formula expressing $\Pi$, the lifetime of the temporal graph and the treewidth of the footprint combined.
\end{toappendix}

Intuitively, we use the static auxiliary graph $H$ to encode the temporal graph $\mathcal{G}$ of size a linear factor of $\tau$ larger than $\mathcal{G}$. Then, using the MSO$_2$ formula of the static problem, we encode the corresponding temporally satisfying reconfiguration problem by adapting the original formula to the auxiliary graph and adding the reconfiguration constraints at every timestep. This gives us a formula which is a factor of $\tau$ longer than the formula encoding the static problem.

\begin{theoremrep}
  \label{theorem:mso}
  Let $\Pi_T$ be a reconfiguration problem such that its static version $\Pi_S$ is expressible by an MSO$_2$ formula $\phi$ expressing the property $\Pi$.
  Let $\mathcal{G}=(G,\lambda)$ be a temporal graph such that the treewidth of $G$ is $tw$. There is an algorithm to determining the maximum/minimum size of a reconfigurable sequence for $\Pi_T$ in $\mathcal{G}$ in $\mathcal{O}(f(tw,\tau,|\phi|) \cdot|G|)$ time.
\end{theoremrep}

\begin{proof}

  Let $H$ be a graph produced by \Cref{const:sliding} on $\mathcal{G}=(G,\lambda)$. Since $\Pi_S$ is expressible in MSO$_2$ logic, there is a predicate $\phi(G_t,X)$ indicating if a subset of vertices $X$ is a set respecting $\Pi$ in $G_t$.
  We optimise the size of this set by making $|X|$ the affine function in Theorem~\ref{thm:EMSO-opt}.
  % cardinality constraint is expressed via an evaluation relation as in \Cref{def:evaluation relation}.
  % Let $\phi$ be an MSO formula such that G
  We construct an MSO$_2$ formula to express a set $X$ such that $X \cap V_t$ respects $\Pi$ in $H[V_t]$ for each $t \in [\tau]$ and such that there is a perfect matching between $X \cap V_t$ and $X \cap V_{t+1}$ for each $t \in [\tau-1]$. Recall that, by Corollary~\ref{cor:check reconfiguration sequence}, a perfect matching implies that we can reconfigure between two sets of vertices.

We introduce the following predicate that given a set of edges $M$, two sets of vertices $X$ and $Y$ and a vertex $v$, $match\_edge$ returns \texttt{true} if and only if
$v$ belongs to $X$,
there is exactly one edge $e\in M$ incident to $v$ and
the other endpoint of $e$ belongs to $Y$.

\begin{eqnarray*}
  match\_edge(M,X,v,Y) = &  v \in X \wedge \exists e, \exists u, (e \in M) \wedge (u \in Y) &\wedge  \\
  & inc(e,v) \wedge inc(e,u) &\wedge  \\
  & \forall e', (e'\in M \wedge inc(e',v)) \Rightarrow e = e'.&
  \end{eqnarray*}

  Notice that given two disjoint vertex sets $X$ and $Y$, a set of edges $M$ forms a perfect matching between $X$ and $Y$ if for every $v$ in $X$ we have the subformula $match\_edge(M,X,v,Y)= \texttt{true}$ and for every $u$ in $Y$ we have $match\_edge(M,Y,u,X) = \texttt{true}$.
  Hence, we can formulate the following predicate that, given two vertex sets $X$ and $Y$ and an edge set $E$ returns \texttt{true} if $E$ contains a perfect matching between $X$ and $Y$.

\begin{eqnarray*}
  Matching(X,Y,E) =
  \exists M, \forall v & &\\
                       & (v\not\in X  \wedge v \not\in Y ) &\vee\\
                       & match\_edge(M,X,v,Y) &\vee \\
                       &match\_edge(M,Y,v,X). &\\
\end{eqnarray*}

Hence, we can encode $\Pi_T$ with an MSO$_2$ formula $\phi'$ as follows:
\begin{eqnarray}
\phi'(X_1,\dots, X_\tau) :=  \bigwedge\limits_{t\leq\tau} \phi(G_t,X_t) \bigwedge\limits_{t<\tau} Matching(X_t,X_{t+1},E_{t,t+1}).  
\end{eqnarray}
We use the affine function $|X_1|+\ldots+|X_{\tau}|$ to maximise/minimise the sets. The matching subformula ensures that the sets have the same cardinality.
The size of the formula $\phi'$ is in $\mathcal{O}(\tau |\phi|)$ and $|H|\leq |G|\cdot \tau$. So, we can conclude by \Cref{theorem:courcelle,thm:EMSO-opt} that $\Pi_T$ is solvable in $\mathcal{O}(f(tw,\tau,|\phi|) \cdot \tau\cdot|G|)$.
\end{proof}
\begin{toappendix}
\noindent Let $G$ be a static graph. The property of a dominating set $D$ and an independent set $I$ can be formulated as an MSO$_2$ formula as follows:
  \[
    \phi_{\text{DS}} (D) :=  \forall v \in V(G), v \in D \vee (\exists u\in D, uv\in E(G)) \tag{{\sc Dominating Set}}
  \]
  \[
    \phi_{\text{IS}} (I):= \forall v \in I, \forall u \in I, uv \not\in E(G) \tag{{\sc Independent Set}}. 
  \]
  %Hence, by \Cref{theorem:mso}, we can deduce the following result for {\sc Temporal Dominating Set Reconfiguration} and {\sc Temporal Independent Set Reconfiguration}.

  Note that these formulae have constant length. For \textsc{Dominating Set} we aim to minimise $|D|$, and we maximise $|I|$ for \textsc{Independent Set}. This gives us the following corollary.
\end{toappendix}
\begin{corollary}
{\sc Temporal Dominating Set Reconfiguration} and {\sc Temporal Independent Set Reconfiguration} are solvable in $\mathcal{O}(f(tw,\tau) \cdot \tau\cdot|G|)$ time.
\end{corollary}

%%% Local Variables:
%%% mode: LaTeX
%%% TeX-master: "main"
%%% End:

\section{Conclusion and future work}
Motivated by both the ability of temporal graphs to model real-world processes and a gap in the theoretical literature, we have defined a general framework for formulating vertex-selection optimisation problems as a temporally satisfying reconfiguration problems, and have described several associated algorithmic tools.  

While hardness results on static vertex selection problems will straightforwardly imply hardness for their corresponding reconfiguration versions, we have described several algorithmic approaches, including an approximation algorithm and several fixed-parameter tractable algorithms.  

Several areas of future work present themselves: first, further investigation of which problems can be solved using our results, or which other temporal parameters are useful here.  
Secondly, because temporal problems are so frequently harder than corresponding static ones, it may be interesting to establish negative results that are stronger than those in the static setting, such as $W[k]$-completeness results that consider the lifetime as a parameter.
Finally, we could explore a more restrictive version of the model, where the number of tokens allowed to move at each time-step is bounded, or there are other restrictions on the speed of change of the vertex set.

\section{Acknowledgements}
Tom Davot and Jessica Enright are supported by EPSRC grant EP/T004878/1. For the purpose of open access, the author(s) has applied a Creative Commons Attribution (CC BY) licence to any Author Accepted Manuscript version arising from this submission. No data were created or analysed in this work.

The authors have no competing interests.

 \bibliographystyle{elsarticle-num} 
\bibliography{biblio}
\end{document}